\documentclass[5p,times,onecolumn,double-spaced]{elsarticle}

\usepackage[utf8]{inputenc}
\usepackage{graphicx}
\usepackage{subcaption}
\usepackage{siunitx}
\usepackage{multirow}
\usepackage{chemformula}
\usepackage{hyperref}
\usepackage{tikz}
    \usetikzlibrary{positioning, shapes, snakes}

\newcommand{\mathendl}{.}

\def\esym {$E_{\rm sym}(\rho)$}
\def\krsn {${\rm ^{86}Kr+^{124}Sn}$}
\def\rpion {$\pi^+/\pi^-$}

\address[thu]{Department of Physics, Tsinghua University, Beijing 100084, China}
\address[ucas]{University of Chinese Academy of Sciences, Beijing 100049, China}
\address[imp]{Institute of Modern Physics, Chinese Academy of Sciences, Lanzhou 730000, China}
\address[cicq]{Collaborative Innovation Center of Quantum Matter, Tsinghua University, Beijing 100084, China}
\address[lzu]{Lanzhou University, Lanzhou 730000, China}

\begin{document}

\title{A CsI hodoscope on CSHINE for Bremsstrahlung $\gamma$-rays in Heavy Ion Reactions}

\author[thu]{Yuhao Qin} 

\author[thu]{Dong Guo} 

\author[thu]{Sheng Xiao} 

 \author[thu]{Yijie Wang}
 
 \author[thu]{Fenhai Guan}
 
 \author[thu]{Xinyue Diao}
 
 \author[thu]{Zhi Qin}

\author[thu]{Dawei Si}    

\author[thu]{Boyuan Zhang}  

\author[thu]{Yaopeng Zhang} 

\author[imp]{Xianglun Wei}

\author[imp]{Herun Yang}

\author[imp]{Peng Ma}

\author[lzu,imp]{Haichuan Zou}

\author[imp]{Tianli Qiu}

\author[imp]{Xinjie Huang}

\author[imp]{Rongjiang Hu}

\author[imp]{Limin Duan}

\author[imp]{Fangfang Duan}

\author[imp]{Qiang Hu}

\author[imp]{Junbing Ma}

\author[imp]{Shiwei Xu}

\author[imp]{Zhen Bai}

\author[imp]{Yanyun Yang}

\author[thu,cicq]{Zhigang Xiao}

\ead{xiaozg@tsinghua.edu.cn}


\date{July 2022}

\begin{abstract}
   Bremsstrahlung $\gamma$ production in heavy ion reactions at Fermi energies carries important physical  information including the nuclear symmetry energy at supra-saturation densities. In order to detect the high energy  Bremsstrahlung $\gamma$ rays, a hodoscope consisting of  15 CsI(Tl) crystal read out by photo multiplier tubes has been built, tested and operated in experiment. The resolution, efficiency and linear response of the units to $\gamma$ rays  have been studied using radioactive source and $({\rm p},\gamma)$  reactions. The inherent energy resolution of  $1.6\%+2\%/E_{\gamma}^{1/2}$ is obtained. Reconstruction method has been established through Geant 4 simulations, reproducing the experimental results where comparison can be made.  Using the reconstruction method developed, the whole efficiency of the hodoscope is about $2.6\times 10^{-4}$ against the $4\pi$ emissions at the target position, exhibiting insignificant dependence on the energy of incident $\gamma$ rays above 20 MeV.  The hodoscope is operated in the experiment of \krsn~ at 25 MeV/u, and a full $\gamma$ energy spectrum up to 80 MeV has been obtained.     
   
\end{abstract}

\maketitle

\section{Introduction}

Heavy ion reactions (HIR) in wide energy range provide a unique way to create  nuclear matter away from the saturation density  in terrestrial laboratory. One of the physical aims of HIR is to study the equation of state (EOS) of nuclear matter, particularly for the matter with larger neutron-to-proton asymmetry, namely the density dependent nuclear symmetry energy \esym. It is a longstanding frontier of common interest of  nuclear physics and astrophysics. Since the discovery of the GW170817 neutron star (NS) merging, the studies of \esym~ has drawn increasing attention for its relevance to the compactness of NS, which influences the amplitude and frequency of gravitational waves \cite{Abbott17,Abbott18,De18,NBZhang2018,NBZhang2019,Xie2019,ALi2018}. 

So far in HIR, a variety of isospin probes have been identified to constrain \esym~ over a wide density range. For instance,  n/p differential flow\cite{yongjiawang15,yongjiawang20}, \rpion~\cite{spirit21} {\it etc.} for suprasaturation densities, and isospin diffusion\cite{tsa04,lba04,clw05}, particle yield ratios \cite{tsa09},  preequilibrium n/p yield ratio \cite{fam06}, angular distribution of $N/Z$ of light clusters  \cite{zhangy07} {\it etc.} for sub-saturation densities. In addition, the static properties of heavy nuclei, like the thickness of neutron skin and the polarizability of $^{208}$Pb  has also been used to  constrain \esym \cite{Tamii2011,ZZ2015}. The current constraint of the \esym~ near $\rho \approx \rho_0$ is consistent among the aforementioned probes.  However, the recent PREX experiment obtained a large neutron skin thickness, leading to a quite stiff \esym~ in tension with the existing results. Despite of the great progress achieved in the last decades, there is still large room to improve the accuracy of the density behavior of \esym, especially at high densities. So the search for further probes of \esym~ is ongoing, for instance, isovector reorientation effect of deuteron scattering \cite{ou15,ou20}, neutron-rich emission from the neck formed in fast fission process \cite{wuqh20}, are also proposed as a sensitive probe of \esym~ near the saturation density, while the isobaric baryonic ratios, $\Sigma^-/\Sigma^+$ and $\Xi^{-}/\Xi^{0}$ are proposed for high densities \cite{yong22}. Combining the GW170817 NS merging event and heavy ion collision experiment, further constraint of \esym~ has been obtained \cite{nature22}. 

Among various probes,  the Bremsstrahlung high-energy $\gamma$ created in heavy ion collisions has been reported to be sensitive to the stiffness of \esym~ and it is advantageous because $\gamma$ rays experience rare final interactions with the medium after it is produced . But the experimental data on this direction is quite scarce. Considering the measurement of  the Bremsstrahlung $\gamma$, which is usually created at the first stage of the collision and the energy can be much higher than the beam energy per nucleon,  the device to measure such high-energy $\gamma$ is not the same as that usually adopted in nuclear spectroscopy studies. The electromagnetic calorimeter is widely used in $e^+e^-$ collider, like BES-III\cite{BESIII-const}, and high energy hadron collider experiment like BaBar\cite{BaBar-EMC}. The high energy $\gamma$ detector of similar type in heavy ion experiment at Fermi energies is also of high interest, for instance, the TAPS experiment\cite{TAPS-IEEE,TAPS-NIMA}. 

In this paper, we report the design, construction and test of a high-energy $\gamma$ hodoscope on the compact spectrometer for heavy ion experiment (CSHINE) running on the radioactive ion beam line at Lanzhou (RIBLL-1). The purpose of this hodoscope is to measure the Bremsstrahlung $\gamma$  in heavy ion reactions at $20-40$ MeV/u incident energy. The paper is organized as following. Section 2 presents the simulations and design of the hodoscope. Section 3 presents the test performance with $\gamma$ source and proton beam. Section 4 presents the experimental performance in beam experiment with the setup of CSHINE. Section 5 is the conclusion. 
 




\section{Design}

\subsection{Mechanics and Materials}
In order to detect the bremsstrahlung photons in heavy ion reactions at Fermi energies, a  $\gamma$ hodoscope is designed and mounted on CSHINE \cite{NIMA-CSHINE}, marked as CSHINE-Gamma hereafter. It aims at detecting the $\gamma$ rays with energy up to about $\SI{80}{\mega\eV}$. Such  high energy photons produce  electromagnetic shower in the detector materials. Thus, sufficient volume of sensitive  material is required to generate and absorb the  shower. CsI(Tl) scintillator  is a kind of high-Z material, less deliquescent and easy to be machined. The radiation length  of CsI(Tl) is $x_0=\SI{8.39}{\gram/{\centi\meter}^2}$, and the Moliere radius is  $R_{\rm M} = \SI{3.531} {\centi\meter}$. Compared to NaI(Tl), although CsI(Tl)  has a slower decay time,  more photons per $\si{\mega\eV}$ can be emitted, making it more efficient in high-energy $\gamma$ ray detection.  Indeed, CsI(Tl) crystals are usually applied in large-scale electromagnetic calorimeter \cite{BESIII-const,BaBar-EMC}.

 CSHINE-Gamma consists of 25 CsI(Tl) detector units, along with  photo multiplier tubes (PMTs)  and readout electronics.  The CsI(Tl) crystals  are manufactured by the Institute of Modern Physics, Chinese Academy of Science. Each unit  has a size of $\SI{70}{\milli\meter}\times\SI{70}{\milli\meter}\times\SI{250}{\milli\meter}$, and one of  the $\SI{70}{\milli\meter}\times\SI{70}{\milli\meter}$ surfaces is the $\gamma$ entrance, corresponding to a transverse dimension of $\sim1.98 R_{\rm M}$, while the length of 250 mm corresponds to  $\sim13.4x_0$. 
 
\begin{figure}[h]
    \centering
    \includegraphics[width=\linewidth]{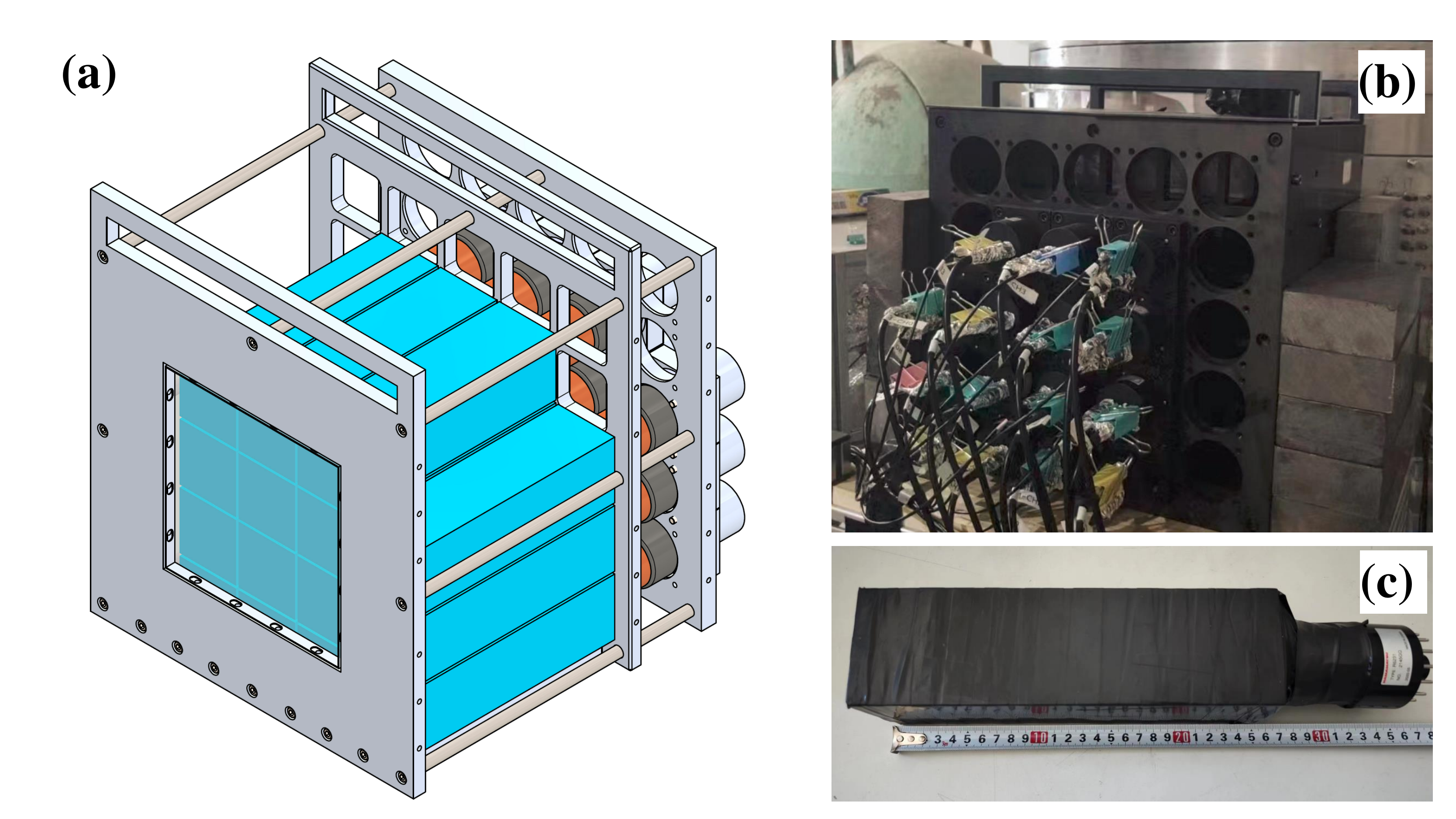}
    \caption{(Color online) Front view of the mechanic frame  (a) and the rear view photograph (b) of the  $\gamma$ hodoscope mounted outside of the CSHINE chamber. (c) is the photo of a single CsI(Tl) scintillator unit coupled to a PMT. }
    \label{fig:design-and-photo}
\end{figure}

 All detector units are mounted on a duralumin frame, forming a $5\times 5$ closely packed hodoscope  (15 units are installed in the current experiment).  Fig. \ref{fig:design-and-photo} (a-b) presents the mechanical design of the hodoscope (front view) and the real photo (rear view) in experiment, respectively. The crystal is wrapped by Teflon tapes for good light reflection, except for the rear surface  coupling to the PMT of Hamamatsu R2631 by BC-630 silicone grease from Saint-Gobain. The wrapped crystal and PMT are then packed by black tapes for light shielding.   Fig.  \ref{fig:design-and-photo} (c) presents the photo of one packed CsI(Tl) unit. The PMT is inserted to Hamamatsu E1198-23  socket, which  distributes the high voltage provided by CAEN N1471 module and extracts the  signal by AC coupling.

According to our simulations, high energy $\gamma$ photons around \SIrange{20}{80}{\mega\eV} directly incident at the unit center should have a fair chance of $>80\%$ to depose $>80\%$ energy within one crystal and its neighboring crystals, as further discussed in the simulation section.


\subsection{Electronics System}

\begin{figure}[h]
    \centering
    \includegraphics[width=0.9\linewidth]{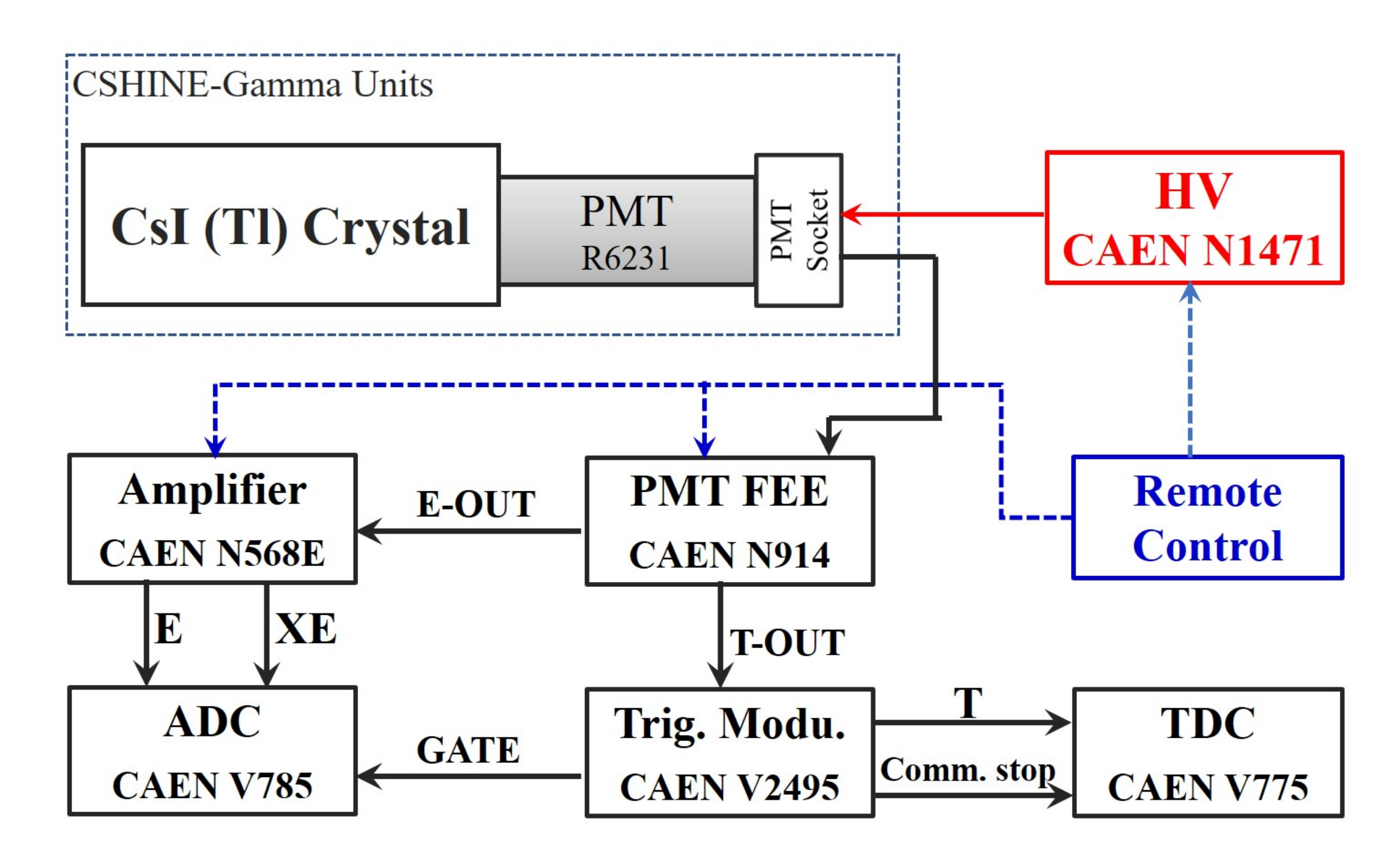}
    \caption{Readout electronic system of CSHINE-Gamma}
    \label{fig:elec}
\end{figure}

The readout electronic system of the $\gamma$ hodoscope is presented in Fig. \ref{fig:elec}.  The  signal from each PMT is  transferred to the front end electronics module CAEN N914 delivering the  amplitude signal and the fast timing signal obtained by leading-edge discrimination. The  negative-polarized amplitude signals from N914 are further transferred to CAEN N568E spectrometer amplifier. In order to record low energy and high energy  $\gamma$ rays, both the OUT and XOUT (with an extra gain by a factor of about 10) signals,  marked by E and XE, respectively,  are digitized by the amplitude-digit converter (ADC) CAEN V785 and saved by the data acquisition (DAQ) system.

The fast time output of CAEN N914 module provides timing and trigger signals. The thresholds of the discriminators are set to a {\it common} level of $\SI{-200}{\milli\volt}$ corresponding to the  $\gamma$ energy around $\SI{1}{\mega\eV}$ in the beam experiment. The rising time and time resolution of the CsI(Tl) units are both around $\sim\SI{40}{\nano\second}$, allowing $\gamma$ photons from different beam pulses to be clearly separated. The timing signals are input to the trigger unit of CAEN V2495, where the whole trigger circuit is constructed based on  FPGA technology, for the details, we refer to \cite{CSHINE-Trigger}. 


\section{Radioactive $\gamma$ source test}

Two types of radioactive sources, \ch{^{60}Co} and \ch{^{232}Th}, are used to test CSHINE-Gamma. The \ch{^{60}Co} source, with a radioactivity of \SI{7.2(6)}{\kilo\becquerel},  provides $\SI{1.173}{\mega\eV}$ and $\SI{1.332}{\mega\eV}$ $\gamma$-ray with $>99\%$ branch ratio. The decay chain of \ch{^{232}Th}  contains \ch{^{208}Tl}  which emits single energy  $\gamma$-ray at $\SI{2.614}{\mega\eV}$.

\begin{figure}[h]
    \centering
    \includegraphics[width=.9\linewidth]{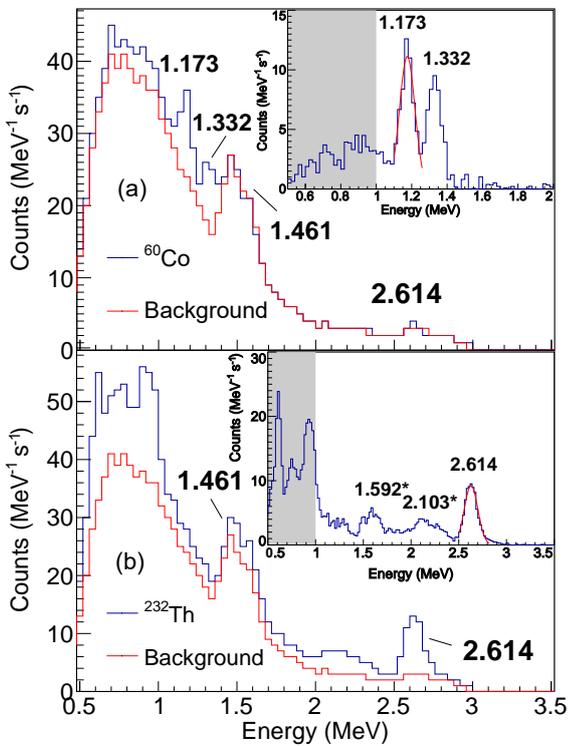}
    \caption{(Color online) \ch{^{60}Co} and \ch{^{232}Th} $\gamma$ spectrum from the CsI(Tl) crystal No. 5.}
    \label{fig:source}
\end{figure}

The \ch{^{60}Co} source is placed at the target location  as in experiment.  The energy spectrum of \ch{^{60}Co} (blue histogram) is shown in Fig. \ref{fig:source}(a). The $\SI{1.173}{\mega\eV}$ and $\SI{1.332}{\mega\eV}$ full energy peaks are shown. Besides, on the background spectrum  accumulated in the same DAQ live time (red histogram),  the $\SI{1.461}{\mega\eV}$ peak from the environmental \ch{^{40}K}  and  the $\SI{2.614}{\mega\eV}$ peak from  the environmental and the self-containing  \ch{^{208}Tl}  are presented in addition to enormous contributions below 1 MeV. By subtracting the background spectrum, the two peaks of the \ch{^{60}Co} sources are clearly distinguished, as shown in the inset of panel (a).  Fitting the full energy peaks with a Gaussian function, the energy resolution of $3.8\% (1\sigma)$  can be extracted for a single CsI(Tl) unit. The signals below 1 MeV will be suppressed by threshold settings in real experiment and are not discussed in this study.

The \ch{^{232}Th} source, placed at the target position too,  is in form of 25 g natural \ch{ThO_2} power contained in a polyethylene bottle, emitting the \SI{2.614}{\MeV} $\gamma$ photons with about \SI{3.2e4}{\becquerel} intensity. Fig.  \ref{fig:source} (b) present the energy spectrum  in\ch{^{232}Th} source test. On the total spectrum (blue histogram), the $\SI{2.614}{\mega\eV}$ full energy peak and the escape peak at $\SI{2.1}{\mega\eV}$ are visible in addition to the strong \ch{^{40}K}  $\SI{1.461}{\mega\eV}$ peak, which is mixed with the escape peak at  $\SI{1.6}{\mega\eV}$. Subtracting the background spectrum (red histogram), the full energy peak  at  $\SI{2.614}{\mega\eV}$, the two escape peaks at  $\SI{2.1}{\mega\eV}$ and $\SI{1.6}{\mega\eV}$ are all presented clearly, as displayed in the inset of panel (b). The energy resolution of $2.6\% (1\sigma)$ is extracted by Gaussian fit for the $\SI{2.614}{\mega\eV}$ full energy peak.

\begin{figure}
    \centering
    \includegraphics[width=.9\linewidth]{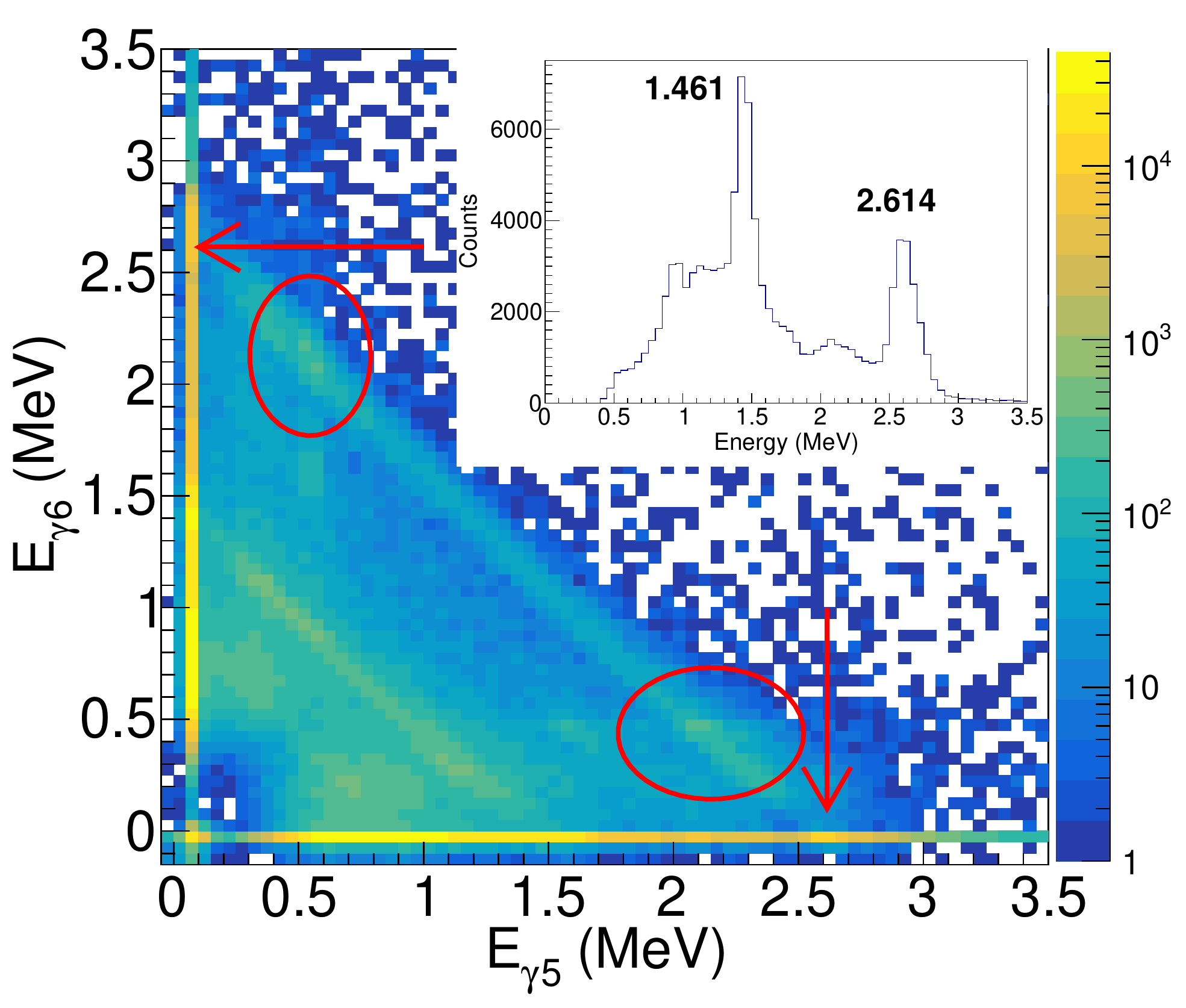}
    \caption{ (Color online) Energy correlation between two neighboring detector units No. 5 and No. 6. }
    \label{fig:Th232-escape}
\end{figure}

In order to see the energy leakage in neighbouring crystals,  Fig. \ref{fig:Th232-escape} presents  the energy correlation between two neighbouring CsI(Tl) units, crystal No. 5 and No. 6 as an example,  in the \ch{^{232}Th} source test. The full energy peak at $\SI{2.614}{\mega\eV}$ in each individual unit is clearly seen, as indicated by the arrows. In addition, there are several bands showing nice anti-correlation. On the band with the sum energy at $\SI{2.614}{\mega\eV}$, the two circled areas refer to the events where two annihilation photons are produced and one of them is caught by the neighboring unit. Summing up the signals above the pedestal level  in the neighbouring units, one can also see a pronounced  full energy peak at  $\SI{2.614}{\mega\eV}$, as depicted in the inset of Fig. \ref{fig:Th232-escape}, indicating the necessity to add back the leaked energies in  neighbouring crystal for high energy $\gamma$ rays.  

\begin{figure}
    \centering
    \includegraphics[width=.9\linewidth]{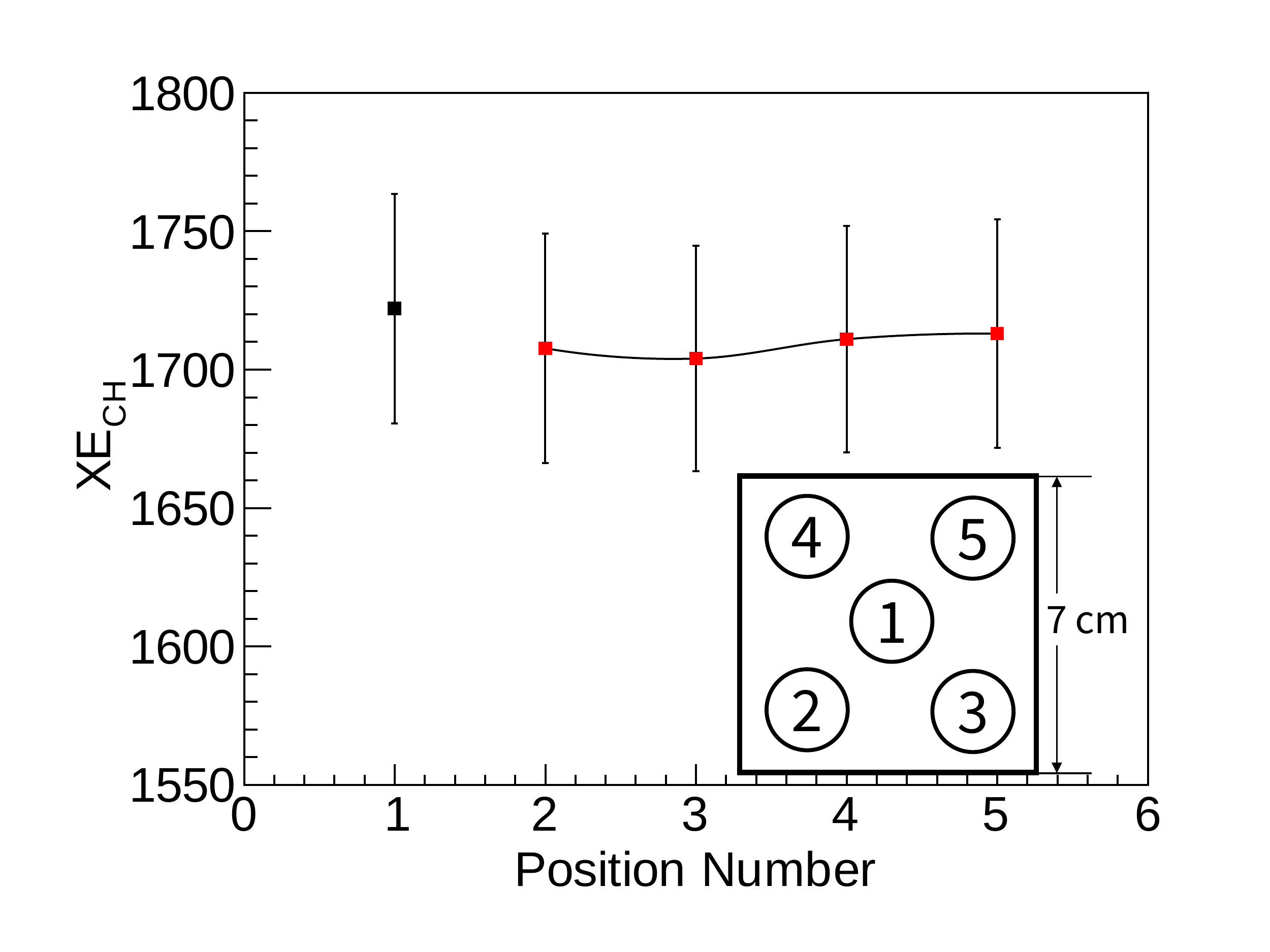}
    \caption{(Color online) Response at different incident position onto the front surface  using a \ch{^{60}Co} source. }
     \label{fig:front-region}
\end{figure}

Because of the purpose  to measure the $\gamma$ rays from the collision on the target,  the response of the crystal to the incident position of the $\gamma$ ray on the front surface has been tested.   The \ch{^{60}Co} source is a good option to test the response because the incident energy is low and probability of energy leakage to neighbouring units is suppressed.  Fig.  \ref{fig:front-region}  presents the ADC channel position of the full energy peak of 1.33 MeV from  \ch{^{60}Co} source as a function of position number, which represents 5 different positions on the front surface as illustrated in the inset. The error bars denote the standard deviation of the peak. It is shown that the peak position differs by less than  $1\%$ varying the incident position for the low energy $\gamma$ rays and the resolution is also constant.  The dependence of the response on the distance between incident position to the readout PMT on the side surface has also been tested, and the variance is less than $10\%$ in most crystals. 

\section{Beam test}

In order to further test the response of the crystals to higher energy $\gamma $ rays, the hodoscope has been tested  by using  high mono-energetic $\gamma$-ray produced in $(\rm{p},\gamma)$ reactions.  The 506 keV proton beam  with about $\SI{1}{\micro\ampere}$ intensity was delivered by  the $2\times1.7$ MV tandem accelerator in China Institute of Atomic Energy (CIAE) \cite{CIAE1,CIAE2}, bombarding on a  natural LiF target. Three mono-energetic $\gamma$-rays can be produced in the reactions \ch{^{19}F(p, $\alpha\gamma$)^{16}O} and \ch{^{7}Li(p, $\gamma$)^{8}Be} \cite{CIAE3}, respectively. The corresponding energies of the $\gamma$ rays are listed  in table \ref{tab:proton_capture}.

\begin{figure}
    \centering
    \includegraphics[width=.9\linewidth]{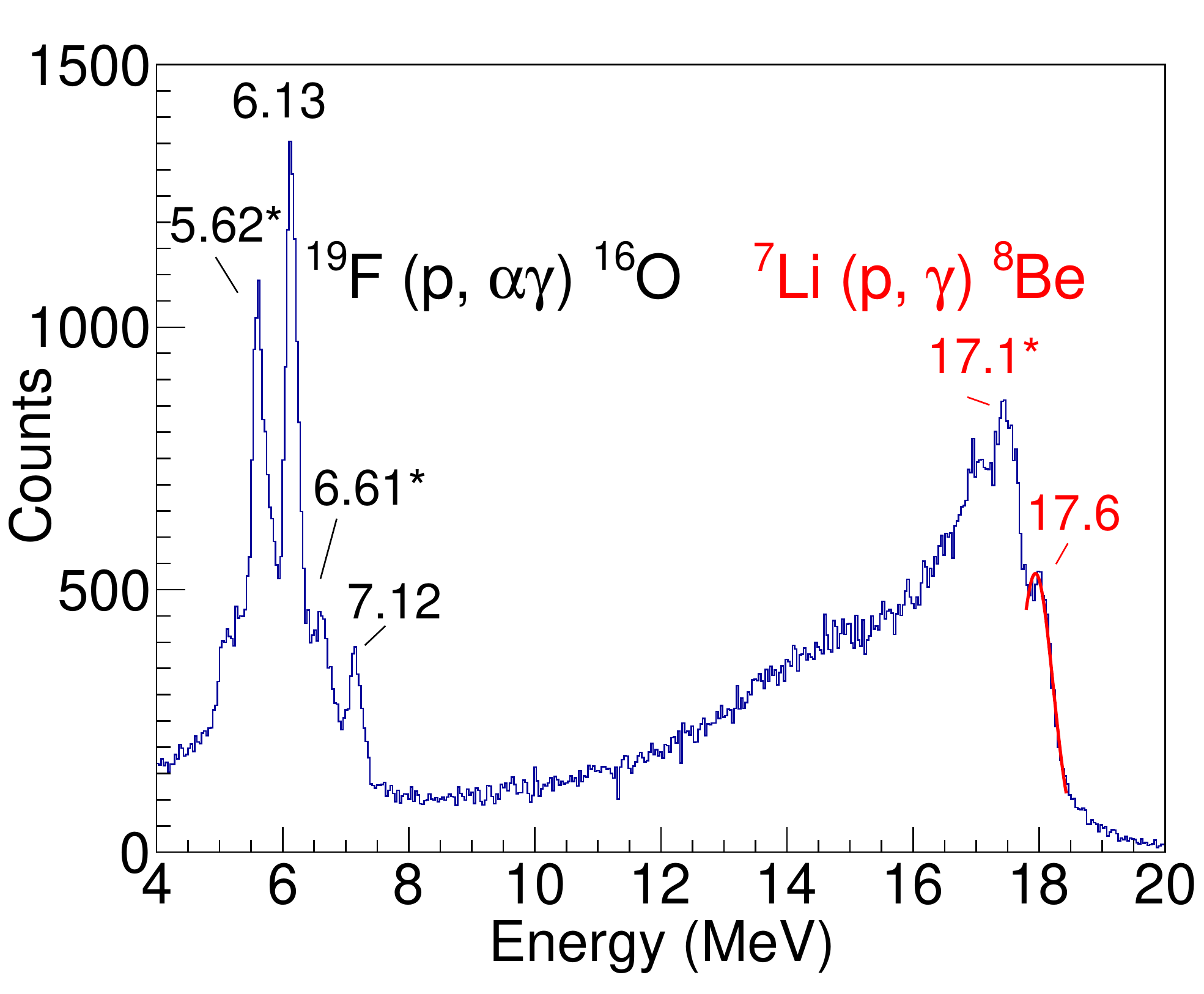}
    \caption{(Color online) The $\gamma$ energy spectrum detected in CsI(Tl) No. 5 in proton capture reactions on \ch{LiF} target at \SI{504}{\kilo\eV} beam energy.}
    \label{fig:beam_test}
\end{figure}

\begin{table}[h]\centering
    \caption{Proton capture reactions on LiF target.}\label{tab:proton_capture}
    \begin{tabular}{ c c c }
        \hline\hline
        Reaction & $E_{\gamma 1}$ ($\si{\mega\eV}$) & $E_{\gamma 2}$ ($\si{\mega\eV}$) \\
        \hline
        {\ch{^{19}F(p, $\alpha\gamma$)^{16}O}}
            & 6.13 
            & 7.12 \\
        \hline
        {\ch{^{7}Li(p, $\gamma$)^{8}Be}}
            & 17.6 
            & / \\
        \hline\hline
    \end{tabular}
\end{table}
Fig.  \ref{fig:beam_test} presents the energy spectrum  detected in the single unit No. 5 as an example. Since the timing information in the beam test was not recorded, the reconstruction by adding up the energy in neighbouring crystals was not conducted due to the lack of timing coincidence.  Nevertheless, various peaks are visible on the calibrated spectrum. On the left part of the spectrum, the 6.13 MeV and 7.12 MeV full energy peaks from \ch{^{19}F(p, $\alpha\gamma$) ^{16}O} are clearly presented. Besides, the  escape peaks missing one 0.511 annihilation $\gamma$ photon are also presented. For the \ch{^{7}Li(p, $\gamma$)^{8}Be} reaction, the 17.6 MeV peak is visible on the high energy tail. Because of the leakage of the energy to neighbouring crystals, the full energy peak is heavily suppressed and the escape components are more pronounced. 

The linearity of the response of the hodoscope can be verified by the mono-energetic peaks in Fig.  \ref{fig:beam_test}. For this purpose, the peaks below 7 MeV are  fitted using linear function. Then one can extrapolate the linear fitting to  17.6 peak, where the residues are checked. For the units where the 17.6 MeV peak is clearly presented and can be fitted unambiguously,the residue distribution ranges from $-0.53$ to 0.74 MeV with a deviation of 0.35 MeV, verifying the linear response of the single CsI(Tl)  to high energy $\gamma$ rays at approximate 2\% uncertainty level.         


\begin{figure}
    \centering
    \includegraphics[width=.9\linewidth]{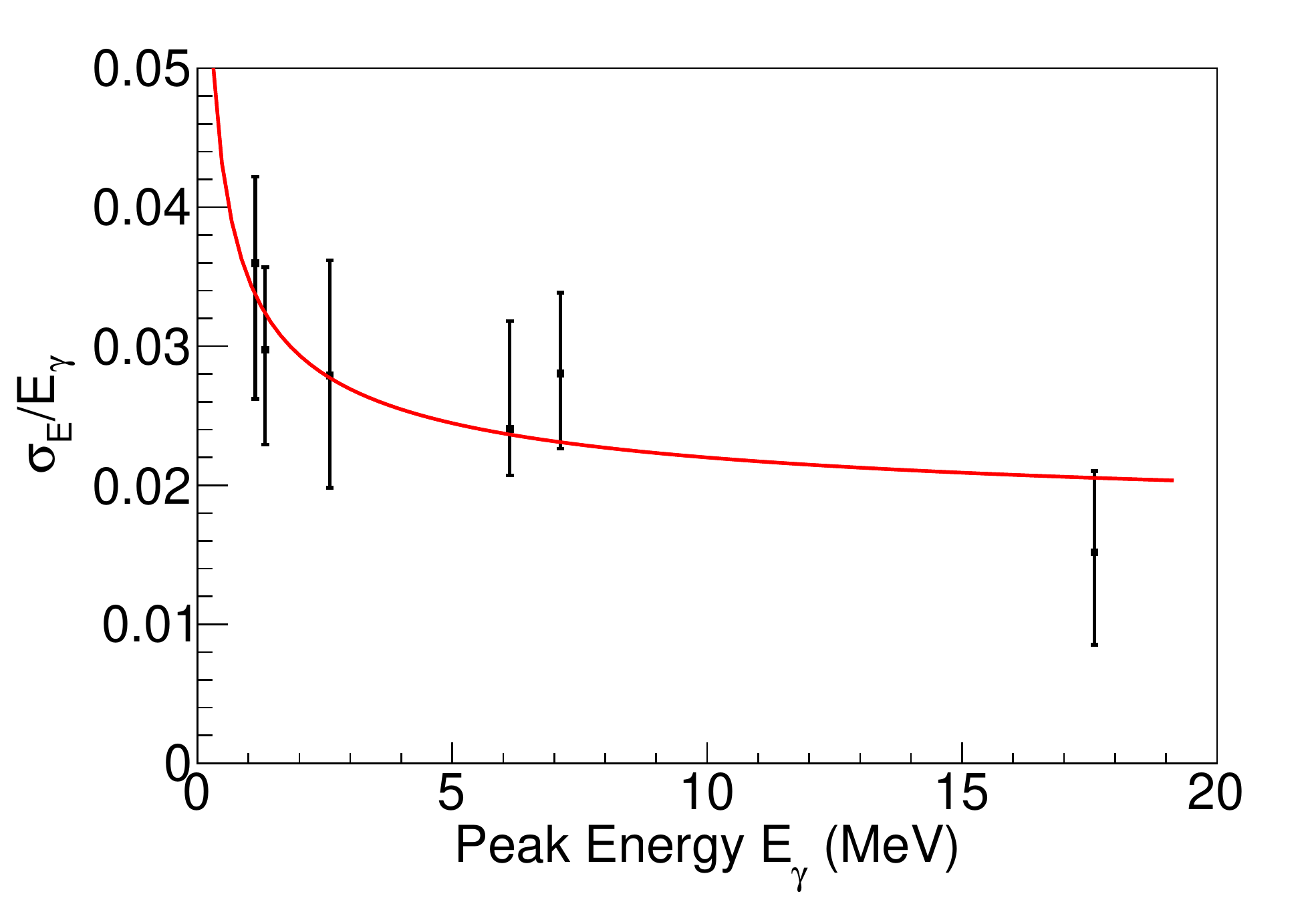}
    \caption{(Color online) Energy resolutions as a function of $\gamma$ energy derived from source test and beam test.}
    \label{fig:resolution}
\end{figure}

Fig. \ref{fig:resolution} presents the averaged energy resolution  defined by the standard deviation of the full energy peak $\sigma_{E}/E_{\gamma}$  of the single CsI(Tl) crystal as a function of incident $\gamma$ energy. The results at 17.6 MeV only contains 7 units where the peak can be separated and fitted.   The error bars on the data points represent the deviation of relative energy resolution  of all units. The solid curve is the fitting result written in $\sigma_{E}/E_{\gamma} = 1.6\%+2\%/E_{\gamma}^{1/2}$.   The average uncertainty level represented by the fitting curve reflects the inherent resolution of a single CsI(Tl) unit, rather than the resolution of high energy $\gamma$ rays in the hodoscope, because the high energy events will be reconstructed from multi Cs(I) units.    

\section{Simulation and Energy Reconstruction}


The Monte-Carlo simulation of CSHINE-Gamma is conducted by Geant4 version 10.5 package\cite{Geant4}. 
The $\gamma$ hodoscope is placed out of the chamber where the target and detectors for charged particles are housed. The source placed at the target is \SI{110}{\centi\metre} to the front surface of CSHINE-Gamma. On the flying path of the $\gamma$-rays from the target to the hodoscope, a square window exists and it is sealed by a duralumin plate with the size of $40\times 40 {\rm cm^2}$ and the thickness of \SI{10}{\milli\metre}  to maintain the vacuum of the chamber. The front plate of the hodoscope frame is  \SI{15}{\milli\metre}, providing mechanic support. A square window of $21\times 21 {\rm cm^2}$  in size is machined on the front frame and covered by a thin duralumin plate of  \SI{1}{\milli\metre} to shield the light  and to reduce the material budget simultaneously for the  central $3\times 3$ units of the hodoscope.  The geometry and the materials of the whole spectrometer, including the CsI(Tl) scintillators and the duralumin shielding structures, are constructed same as in the experiment. 

Two types of simulations have been conducted with Geant4. One is to simulate  the response of $\gamma$ rays from \ch{^{60}Co} source in comparison to experimental test. The other is to simulate the response for  $\gamma$ rays with continuous energy ranging from  \SI{1}{\MeV}  to \SI{80}{\MeV}.

\begin{figure}[h]
    \centering
    \includegraphics[width=.8\linewidth]{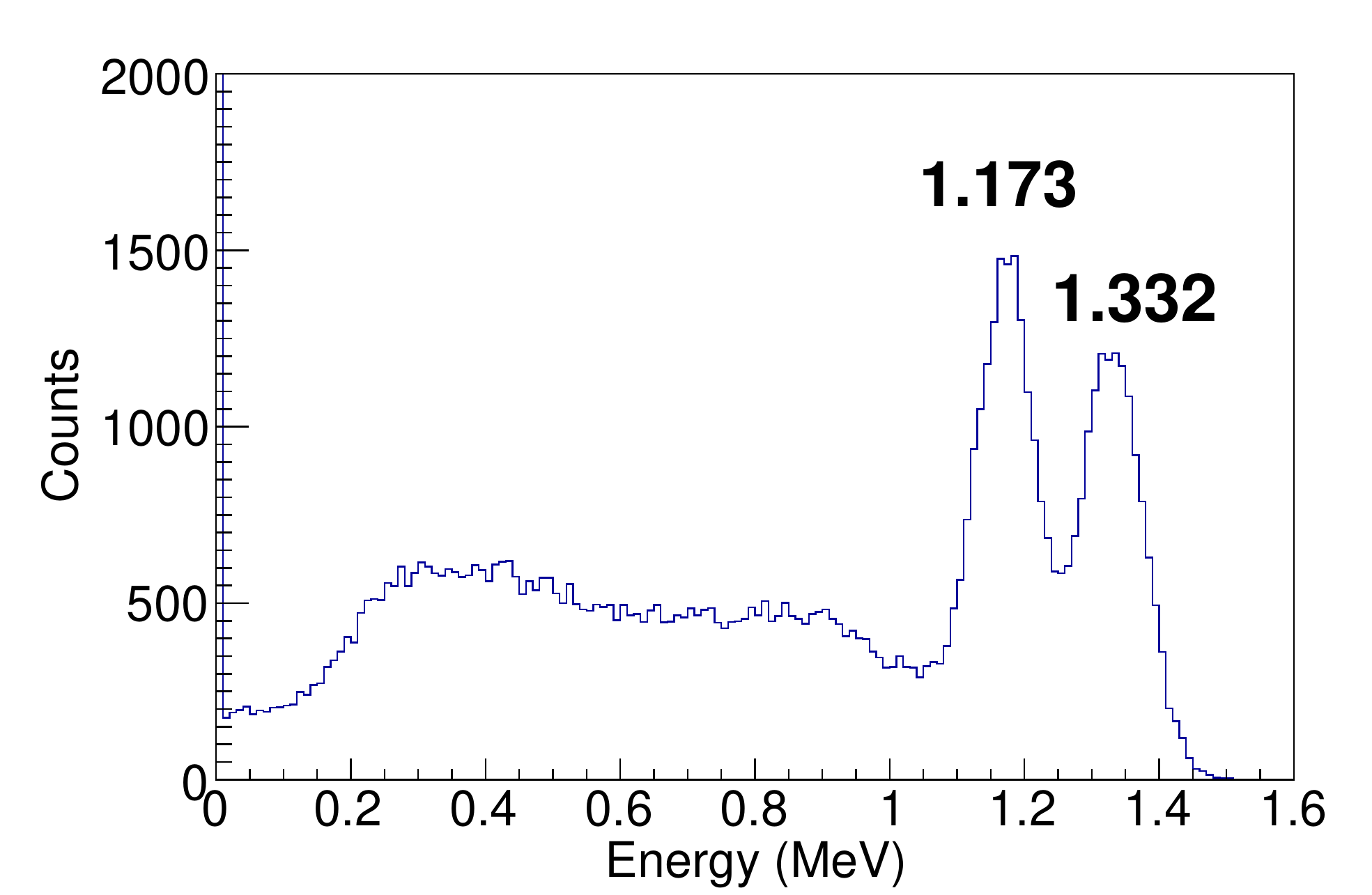}
    \caption{(Color onlie) Simulated \ch{^{60}Co} $\gamma$ spectrum with Geant4.}
    \label{fig:h_sim_Co}
\end{figure}

For the simulation of \ch{^{60}Co},  the two $\gamma$ photons at \SI{1.173}{\MeV} and \SI{1.332}{\MeV} are simulated separately according to the source activity, which is low and unlikely cause pileup on the energy spectra. The energy resolution of  $4\%/{E_\gamma^{1/2}}$ is folded in the simulation. Fig. \ref{fig:h_sim_Co} presents the energy deposit of \ch{^{60}Co} $\gamma$ in one of the detector units, and the two full energy peaks from \ch{^{60}Co} are distinguished.  

\begin{figure}[h]
    \centering
    \includegraphics[width=.8\linewidth]{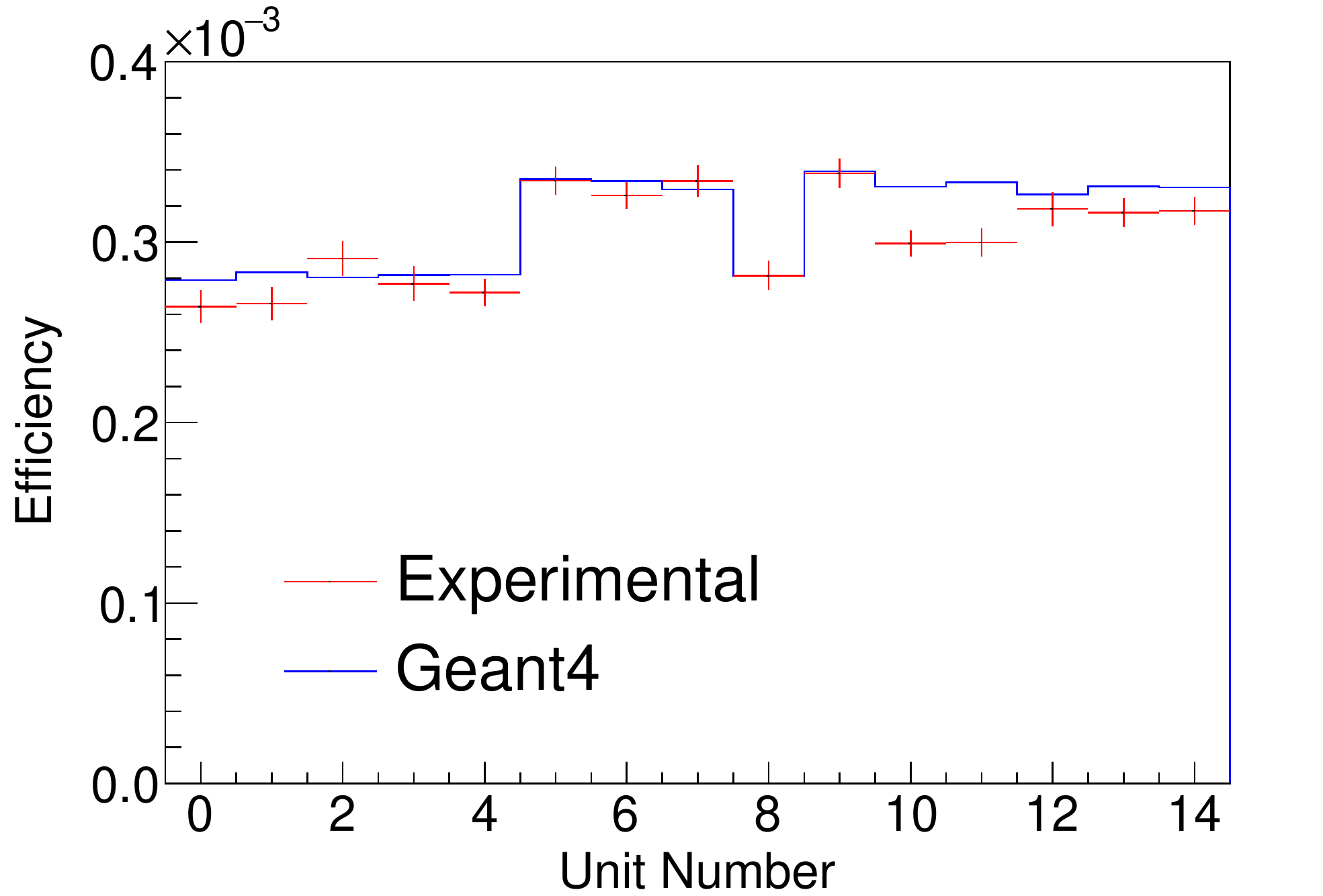}
    \caption{(color online) \ch{^{60}Co} efficiency obtained by source test(red cross) and by Geant4 simulations (blue histogram).  }
    \label{fig:c_eff_Co}
\end{figure}

The efficiency can be verified by the simulation of \ch{^{60}Co}.  Because of the broadening to the full energy peaks, the  efficiency of a single unit is computed by counting the events above 1 MeV out of the total number of incident $\gamma$ rays from \ch{^{60}Co} source.  In real test, due to the background contribution, referring to Fig. \ref{fig:source}(a) for instance, one has to subtract the background  and correct the dead time loss before deriving the efficiency $\eta_{\rm ^{60}Co}$. For clearness, it  can be written as

\begin{equation}
    \eta_{\ch{^{60}Co}} = \frac1{2A} \cdot \left(
        \frac{N_{\ch{^{60}Co}}}{T^{\mathrm{liv}}_{\ch{^{60}Co}}}
        -\frac{N_{\mathrm{bkg}}}{T^{\mathrm{liv}}_{\mathrm{bkg}}}
        \right)
    \mathendl\label{eqn:eff_Co}
\end{equation}
where A is the  radioactivity and  $T^{\mathrm{living}}$ is the living time. $N_{^{60}\rm Co}$ and $N_{\rm bkg}$  are the counts measured in the corresponding living time $T^{\rm liv}_{\rm ^{60}Co}$ and $T^{\rm liv}_{\rm bkg}$, respectively. The factor $1/2$ is presented because of the fact that 2 photons are emitted in each decay of \ch{^{60}Co}.  Fig. \ref{fig:c_eff_Co} presents  the efficiency given by Geant 4 simulations in comparison with the source test for all the 15 units.  It is shown that for most units, the experimental efficiency is reproduced by the Geant4 simulations within 5\%, with two exceptions at No. 10 and 11, for which the energy resolution is slightly worse. The $3\times3$ units of No. 5, 6, 7, 9, 10 to 14 have a higher efficiency because the covering duralumin plate in front of  these units is thinner (1 mm).  From this comparison, one can assure that the efficiency simulated by Geant4  can be applied in the high energy region.

\begin{figure}[h]
    \centering
    \includegraphics[width=.8\linewidth]{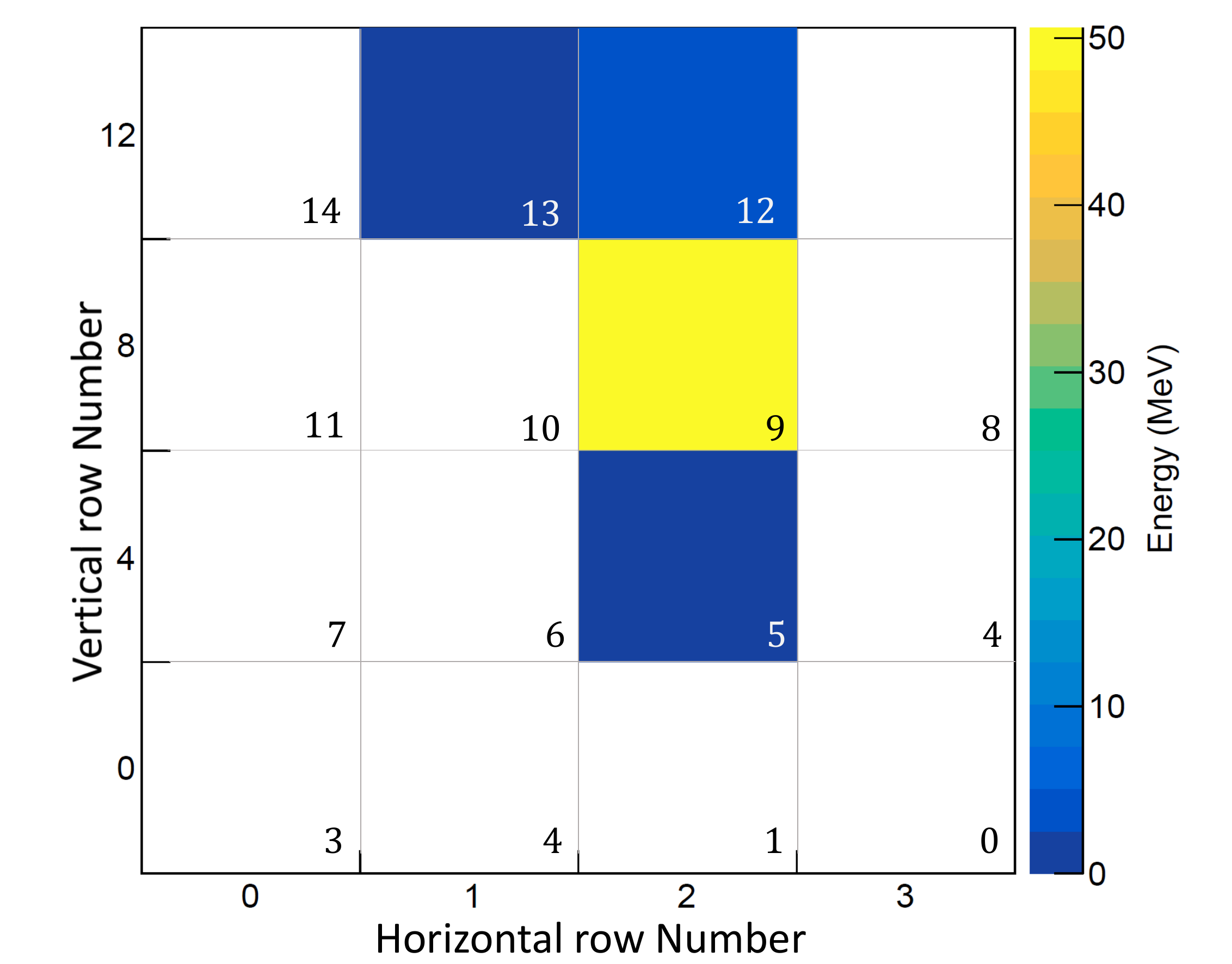}
    \caption{(Color online) Front view of the event center and neighboring crystals. Most of the $\gamma$ energy are collected by the event center, which is No. 9 in this event.}
    \label{fig:reconstruction}
\end{figure}

  Next, the efficiency curve up to high energy can be simulated using the same simulation packages. As the transverse size of one unit is approximately  $2r_{\rm M}$,  the $\gamma$ photons with  energy of tens MeV have a certain probability to leak some energies to the neighboring units via the shower photons or electrons. Thus, one must collect the leaked energy in the crystals surrounding to the incident one. Fig. \ref{fig:reconstruction} presents an event display for a high-energy incident $\gamma$ ray in the simulation. The color represents the energy deposit. The crystal containing  the highest energy deposit is marked as the ``event center, $I_{\rm ec}$'', which is No. 9 in this example. In the current setup, if the event center is one of the 4 central crystals, namely $I_{\rm ec}=5$, 6, 9 or 10 in the hodoscope, the event center is surrounded by 8 other crystals (7 in the case that No. 9 is the event center) and the leaked energy can be added back since the total transverse dimension of the $3\times 3$ pack is much larger than  $2r_{\rm M}$. This type of event  is considered here as the ``best event''  for reconstruction in both simulation and experiment in this paper.

\begin{figure}[h]
    \centering
    \includegraphics[width=.8\linewidth]{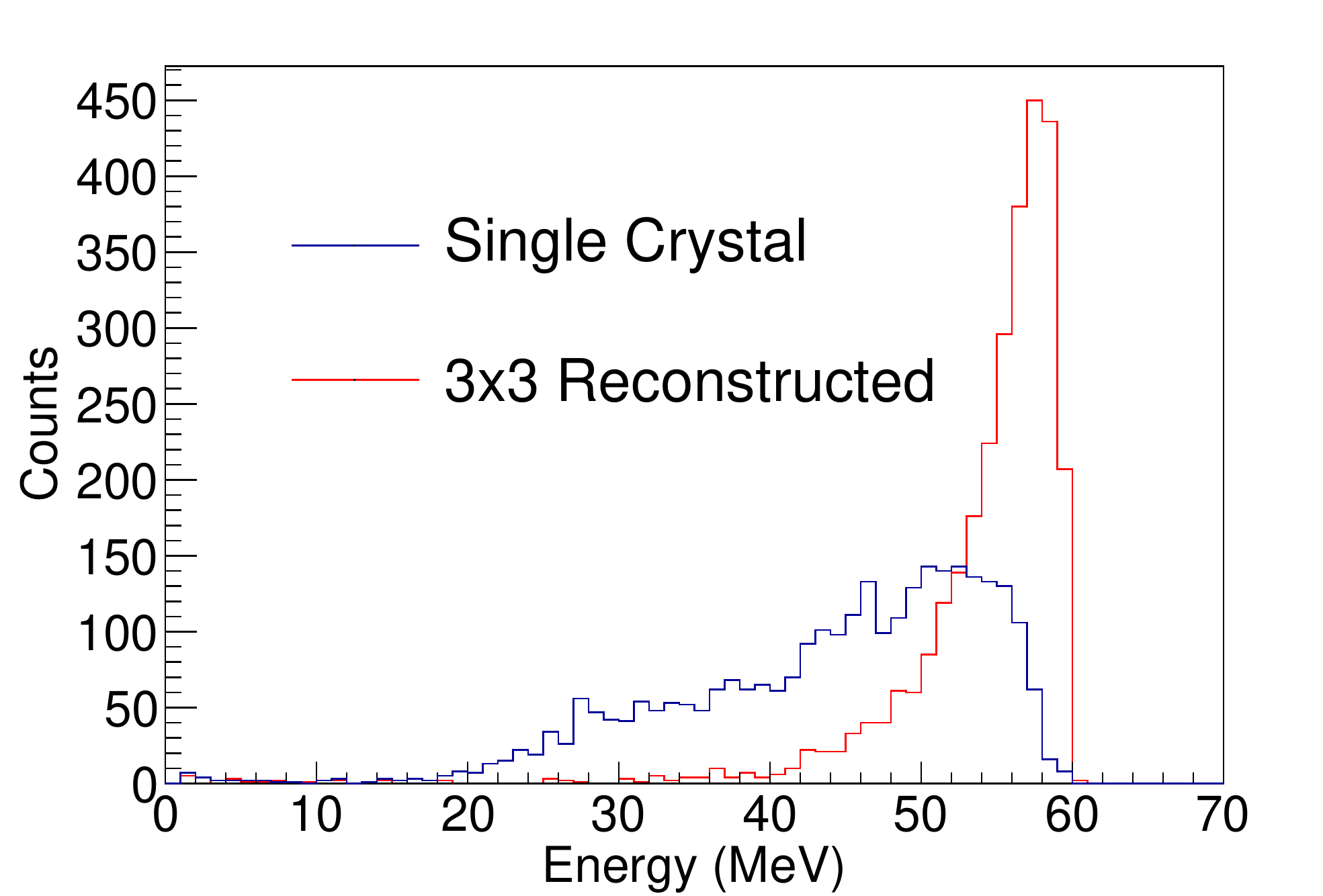}
    \caption{(Color online) Simulated energy deposit of \SI{60}{\MeV} $\gamma$,  before (blue) and after (red)  adding back the neighbouring crystals}\label{fig:h_sim_beam}
\end{figure}

Fig. \ref{fig:h_sim_beam}  presents the simulated spectrum of the ``best event''  for the \SI{60}{\MeV} single-energy photons by counting the signal in the incident crystal only (blue histogram) and the signals in the $3 \times 3$ configurations (red histogram), respectively. It is shown that due to the energy leakage, the shape of the spectrum is  distorted and the full energy peak is suppressed if only the signal event center is counted. By adding back the signals from the surrounding units, on the other hand, the full energy peak at 60 MeV is largely restored, approximately 80\% events contains 80\% of the total incident energy.   

\begin{figure}[h]
    \centering
    \includegraphics[width=.8\linewidth]{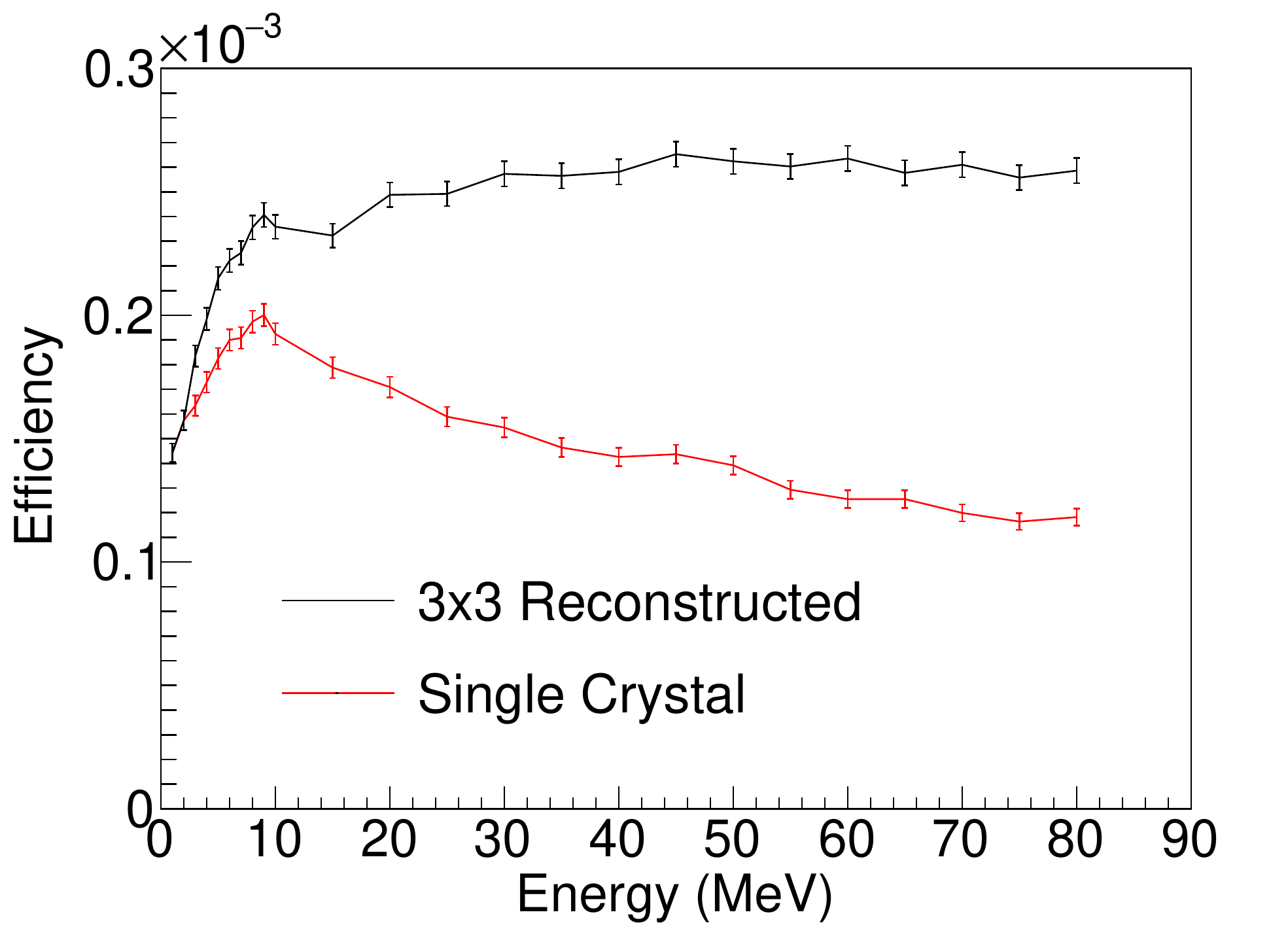}
    \caption{(Color online) Simulated $\gamma$ full energy efficiency (geometry taken into count.)}
    \label{fig:c_eff_sim_beam}
\end{figure}

Fig. \ref{fig:c_eff_sim_beam} further presents the simulated efficiency as a function of the energy of the incident $\gamma$ rays from \SIrange{1}{80}{\MeV},  with and without adding back energy deposit in the neighbouring crystals, respectively. Above 10 MeV,  without counting the neighbouring crystals, the efficiency decreases monotonically because of the leakage. While counting the neighbouring crystals, the efficiency distribution keeps constant above 10 MeV. Here the geometry acceptance is taken into account. The efficiency loss below 10 MeV is due to the absorption of the $\gamma$ rays in the materials on the flight path from the target to the hodoscope.  Worth mentioning, if the event center is on the outer ring of the hodoscope, other than from the central 4 crystals,  the probability that leakage energy is undetected increases. The reconstruction will be more complicated and is  under investigation.

\section{Experiments and Performance}

\begin{figure}[h]
    \centering
    \includegraphics[width=.9\linewidth]{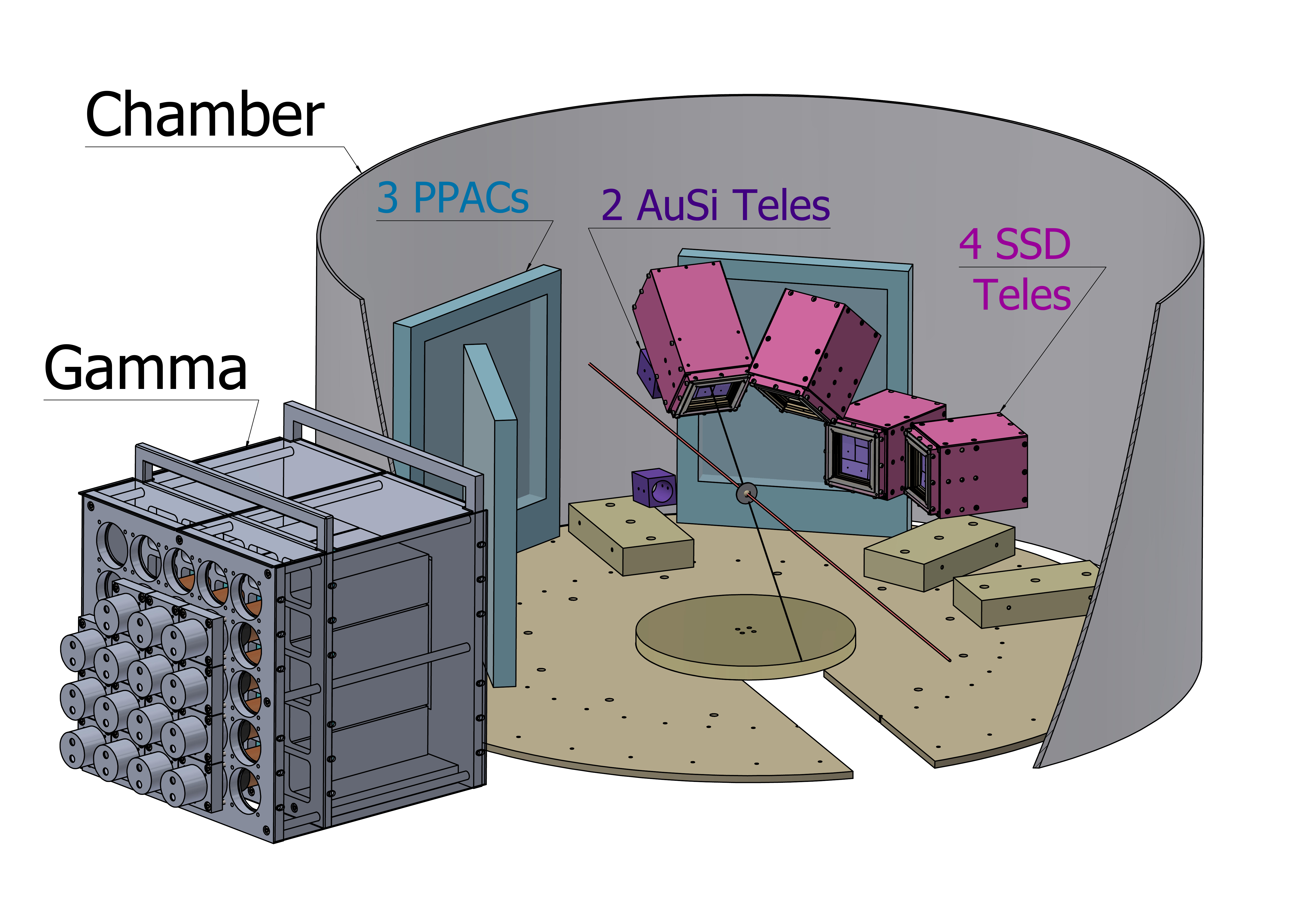}
    \caption{ (Color online) Detector setup for the  KrSn2022 experiment.}
    \label{fig:Krsn2022-setup}
\end{figure}

Finally, we present the performance of CSHINE-Gamma in the experiment campaign of  $^{86}$Kr+$^{124}$Sn at 25 MeV/u ( marked by KrSn2022), performed  at the final forcal plane of the radioactive ion beam line at Lanzhou (RIBLL1) in 2022. The \ch{^{86}Kr} beam  was delivered by the cyclotron of the heavy ion research facility at Lanzhou (HIRFL), bombarding on a \ch{^{124}Sn} target with the thickness of \SI{1}{\milli\gram/\centi\metre^2}. The detector setup of this experiment is shown in Fig. \ref{fig:Krsn2022-setup}. In the current phase,  4 movable SSDTs have been installed covering the polar angle  $\SI{20}{\degree}<\theta<\SI{100}{\degree}$ to measure the charged particles of $Z\leq6$. The coincident fission fragments are measured by the  PPACs installed at 40$^\circ$, 90$^\circ$ and $-40^\circ$  with respect to the beam, respectively. Here the $``-"$ sign means on the other side of the beam. For the performance and previous physical results on isospin chronology obtained by the SSDTs, one can refer to  \cite{NST-CSHINE,NIMA-CSHINE,PLB-WYJ}.  For the performance of the PPACs and the reconstruction of the fission events following the heavy ion reactions, one can refer to  \cite{NST-Diao}. Two Si(Au)-CsI telescopes are mounted to monitor  the elastic scattering  at $\theta_{\rm Lab}= \SI{15}{\degree}$. Besides,  CSHINE-Gamma is mounted outside of the target chamber at $\theta_{\rm Lab}= \SI{100}{\degree}$, being 110 cm to the target. The side surface of CSHINE-Gamma was further shielded by lead bricks.  

The FPGA-based trigger circuit are implemented using the V2495 module. For the details of the trigger system, one can refer to \cite{CSHINE-Trigger}. For most of the beam time, CSHINE-Gamma is not added to the main trigger logics for two reasons. 1) Because the CsI(Tl) crystals are slow detectors, and 2) at the RIBLL1 experimental site, CSHINE-Gamma has single hit rate of $>\SI{1e4}{\second^{-1}}$ when the beam is on. It is  higher than the maximum event rate the DAQ system can handle, and may cause a malfunction randomly.  Thus, in offline analysis, we search the $\gamma$ signals with correct timing information in the events triggered by two fission fragments or by two light charged particles produced in the reactions on the target.  

\subsection{Calibration of CSHINE-Gamma}

  Since both low gain (E) and high gain (XE) outputs are recorded  to cover the wide energy range, the energy responses of both E and XE have to be calibrated consistently  for each individual channel. The calibration procedure consists of three  steps.  i) using experimental data to  establish the linear correspondence between E and XE, which depends only on the electronics,  ii) using \ch{^{60}Co} and \ch{^{232}Th} sources to  convert XE channels to \si{\MeV}, and iii)  to establish the relationship between $E_{\gamma}$ and the ADC channels in the whole energy range by assuming the linear response of CsI unit to $\gamma$ energy.

\begin{figure}[h]
    \centering
    \includegraphics[width=.8\linewidth]{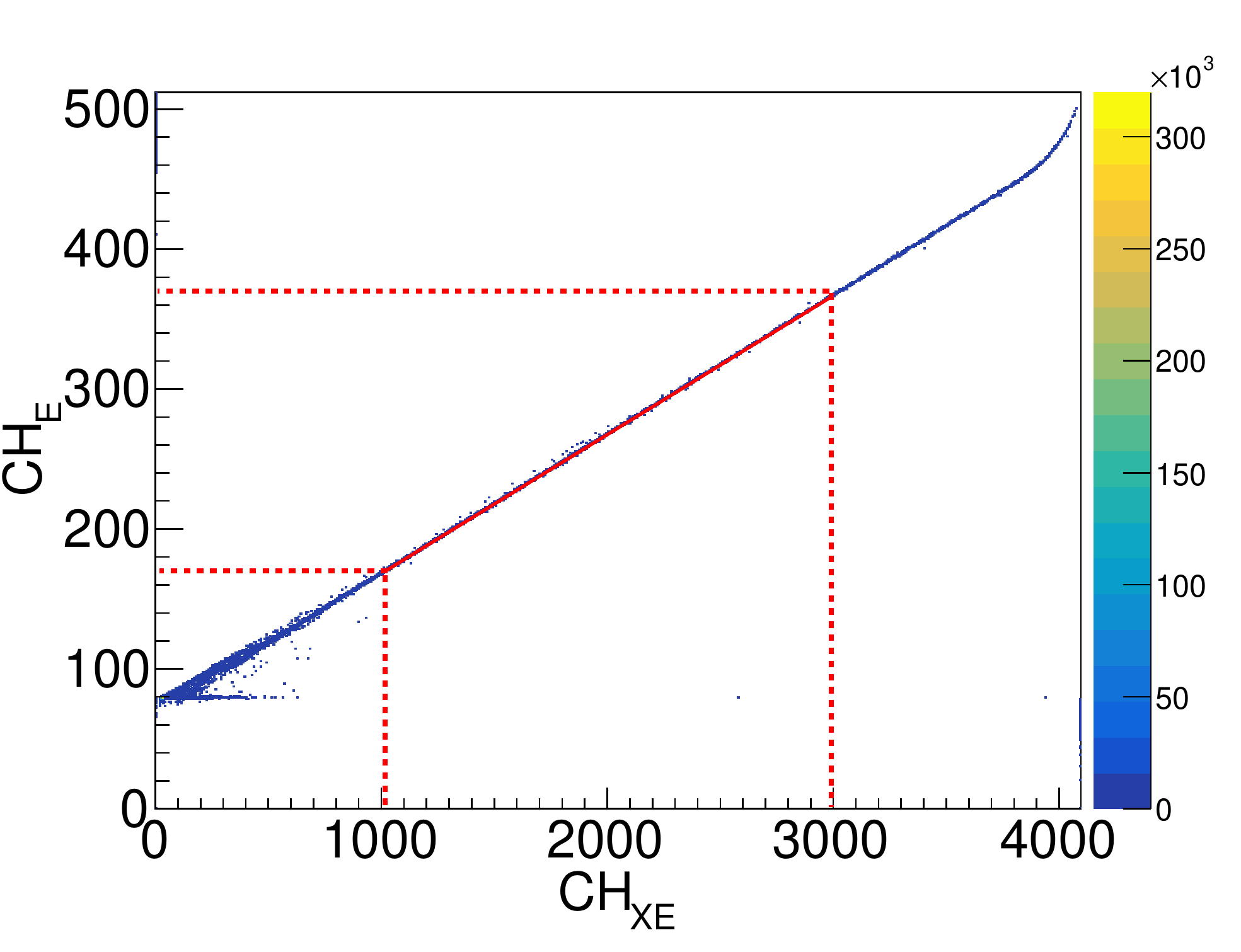}
    \caption{ (color online) ADC E-XE  correlation for one crystal.}
    \label{fig:E-XE}
\end{figure}

First, the linear correspondence between E and XE is established using inclusive $\gamma$ events.  Fig. \ref{fig:E-XE} presents the linear correlation of one unit for example. It can be seen that in a wide range, the E and XE channels keep a good linear relation written as

\begin{equation}
    \mathrm{CH_{XE}}=\alpha \mathrm{CH_E} + \beta
\end{equation}

where $\mathrm{CH_{XE}}$ and $\mathrm{CH_E}$ are the ADC channels of the XE and E output, respectively. $\alpha$ and $\beta$ are the linear coefficients  obtained from  Fig. \ref{fig:E-XE} in the range with good linearity, i.e.,  $1000<\mathrm{CH_{XE}}<3000$.  Once it is determined, although XE output losses its function with $\mathrm{CH_{XE}}>3500$, the above relationship is valid in the whole range of  $\mathrm{CH_E}$.  

\begin{figure}[h]
    \centering
    \includegraphics[width=.75\linewidth]{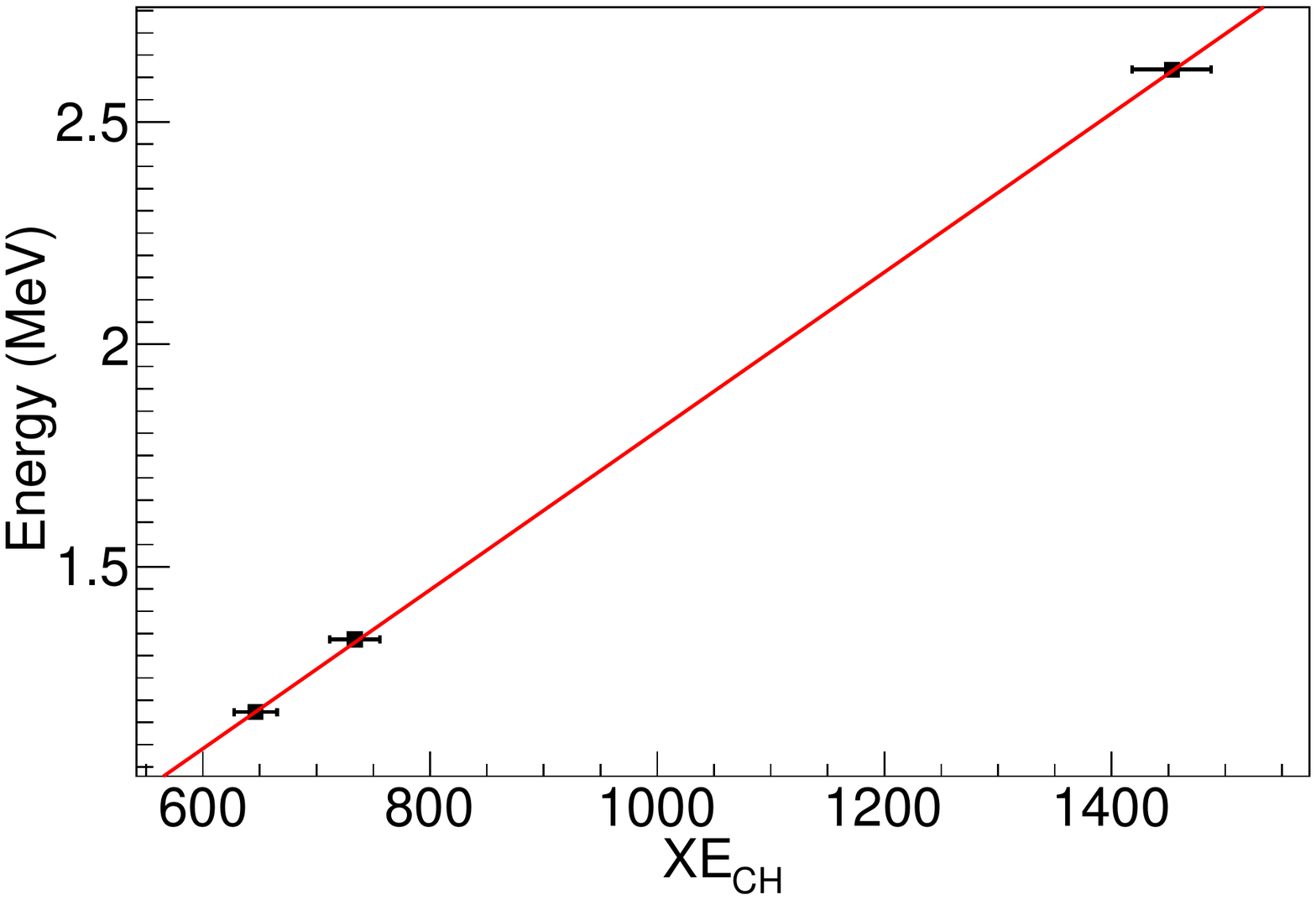}
    \caption{(Color online) Radioactive source calibration for one XE channel}
    \label{fig:XE-source}
\end{figure}

Then, the  source calibration is done off-beam using the \ch{^{60}Co} source and the  \ch{^{232}Th} source.  The three types of $\gamma$ rays, 1.173, 1.332 and \SI{2.614}{\MeV},   can be used to calibrate the energy response of XE.  Fig. \ref{fig:XE-source} shows the correspondence of the three energies and the ADC channels. The linear relation between $\rm CH_{XE}$ and  energy $E_{\gamma}$ in \si{\MeV} can be written in 

\begin{equation}
    E_{\gamma}=a \mathrm{CH_{XE}} + b
\end{equation}

Last, according to the beam test using proton-induced reactions, it is reliable to assume that the response of the CsI(Tl) crystal to the energy deposit is linear. Thus, combining formula (2) and (3), one can measure the high energy incident $\gamma$ rays using the E output at the channel $\rm CH_E$ by 

\begin{equation}
 E_{\gamma}=a \alpha \mathrm{CH_E} + a \beta   + b
\end{equation}

Table \ref{tab:coeff} presents all the linear coefficients of the 15 units in the beam experiment KrSn2022. It is shown that the main uncertainty are contributed by the inherent resolution of the CsI(Tl), manifesting itself in the uncertainties of $a$ and $b$  from the source test. For most of the crystals, the uncertainties of coefficient $a$ are about $5\%$, in accordance with the results shown in Fig. \ref{fig:resolution}, except for the two units  No. 11 and 10, which have  worse  resolution at about $7-10\%$.   

\begin{table}[h]\centering
    \caption{Linear coefficients of each unit\label{tab:coeff}}
    \begin{tabular}{ c | c c c c }
        \hline\hline
        Unit&	$\alpha$&	$\beta$&	$a$&	$b$\\
        \hline
        0&	0.09988(1)&	94.65(1)&	0.00195(10)&	-0.15(12)\\
        1&	0.09967(1)&	70.49(3)&	0.00180(10)&	-0.10(8)\\
        2&	0.09957(1)&	85.93(3)&	0.00190(10)&	-0.16(9)\\
        3&	0.09980(1)&	56.18(1)&	0.00189(8)&	-0.09(7)\\
        4&	0.09948(1)&	77.35(1)&	0.00186(9)&	-0.14(8)\\
        5&	0.09930(4)&	68.82(7)&	0.00179(9)&	0.02(7)\\
        6&	0.10014(4)&	74.27(7)&	0.00180(8)&	-0.19(8)\\
        7&	0.09924(1)&	72.60(1)&	0.00177(9)&	-0.05(8)\\
        8&	0.09970(2)&	91.27(3)&	0.00198(9)&	-0.18(7)\\
        9&	0.09977(1)&	70.84(2)&	0.00180(12)&	-0.08(10)\\
        10&	0.09954(1)&	70.62(1)&	0.00210(14)&	-0.14(10)\\
        11&	0.09974(3)&	86.81(5)&	0.0020(2)&	-0.15(17)\\
        12&	0.10007(2)&	82.09(4)&	0.00163(7)&	-0.12(7)\\
        13&	0.10009(3)&	75.44(5)&	0.00167(10)&	-0.13(10)\\
        14&	0.09984(1)&	109.27(1)&	0.00191(10)&	-0.13(9)\\
        \hline\hline
    \end{tabular}
\end{table}
 
\subsection{Beam Experiment Performance}

During the experiment, totally $10^6$ $\gamma$ events are detected in coincidence with the reactions on the target, identified by two fission fragments or two LCPs. The energy and time information for the $\gamma$ events are detected.

The timing signal of the each  $\gamma$ hodoscope unit is obtained by leading edge discrimination in CAEN N914 modules, thus the  time-amplitude (T-A) correction is required because of the time walk. The  timing of the incident $\gamma$ ray,  $t_\gamma$, is written as \cite{NST-Diao}:

\begin{equation}
    t_\gamma = t_{\mathrm{det}} - t_{\mathrm{RF}} - t_{\mathrm{corr}}(\mathrm{A}_\gamma)
\end{equation}
where $t_{\mathrm{det}}$ is the TDC channel position in unit of ns,  $t_{\mathrm{RF}}$ is the calibrated  radio frequency (RF) from the accelerator. $t_{\mathrm{corr}}(A_{\gamma})$ is a T-A correction as a function of the amplitude $A_{\gamma}$,  where $\mathrm{A}_\gamma$ is the ADC E signal of the same crystal. Assuming $t_{\mathrm{corr}}$ is inversely proportional to the amplitude, it is then written as

\begin{equation}
    t_{\mathrm{corr}}(\mathrm{A}_\gamma) = 
        C_t / (\mathrm{A}_\gamma - A_{\rm base})
\end{equation}
where $C_t$ is a time constant and $A_{\rm base}$ is the ADC baseline constant of this crystal. After the T-A correction, the distribution of $ t_\gamma$ is independent on the amplitude for each individual crystal. And then, the timing information can be used to establish the coincidence of different crystals that are fired by the same incident high energy $\gamma$ ray. 

\begin{figure}[h]
    \centering
    \includegraphics[width=.9\linewidth]{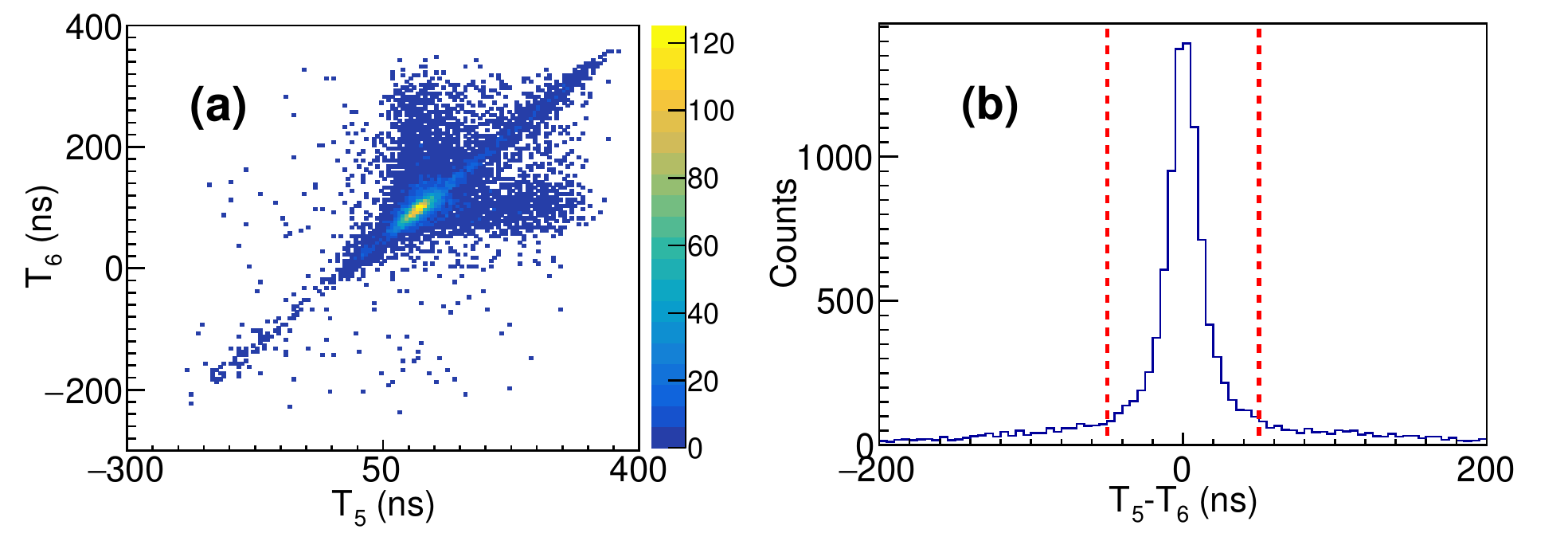}
    \caption{(Color online) Timing correlation for two neighboring crystals}
    \label{fig:h_timing_CH5_CH6}
\end{figure}

Fig. \ref{fig:h_timing_CH5_CH6} (a) presents the timing correlation of two neighboring crystals, No. 5 and 6 again for example,  in coincidence with the fission event requiring the multiplicity of fragments in PPAC is $M_{\rm ppac}=2$. For the majority of the events, it is shown that the timing signals of these two crystals are correlated, indicating these two $\gamma$ rays are from a same event. In order to further exclude that the correlated signals are from two separate $\gamma$ rays with lower energies, we investigated the time signal correlation from two crystals which are not neighbouring, and such correlation is suppressed over an order of magnitude. It suggests that the two correlated signals in No. 5 and No. 6 are unlikely from two separate $\gamma$ rays, otherwise because these events shall also be similarly pronounced in the crystals separated far away.  Panel (b) presents the time difference of the two crystals in (a). The cuts on the time difference $|\Delta  T_{\rm neib}|< 50 ~{\rm ns}$ of two neighboring units is then applied in the energy reconstruction of the $\gamma$ rays.

Finally, in order to exclude the penetrating cosmic ray muons, which leaves large signal in the CsI crystals,  we checked the spatial spreading of signals in each event. The spatial spreading of the signals on horizontal (x) and vertical (y) directions are defined as 

\begin{equation}
    \delta x = E_{\rm i} {|x_{\rm i}-\bar{x}|}/E_{\rm tot} 
\end{equation}
\begin{equation}
    \delta y = E_{\rm i} {|y_{\rm i}-\bar{y}|}/E_{\rm tot}
\end{equation}
where $E_{\rm i}$, $x_{\rm i}$  and $y_{\rm i}$ are the energy, horizontal and vertical position of crystal $i$, respectively. $E_{\rm tot}$ is the total energy.   $\bar{x}$ and  $\bar{y}$ are the weighted center of the incident $\gamma$  horizontally and vertically, respectively. 

\begin{figure}[h]
    \centering
    \includegraphics[width=.9\linewidth]{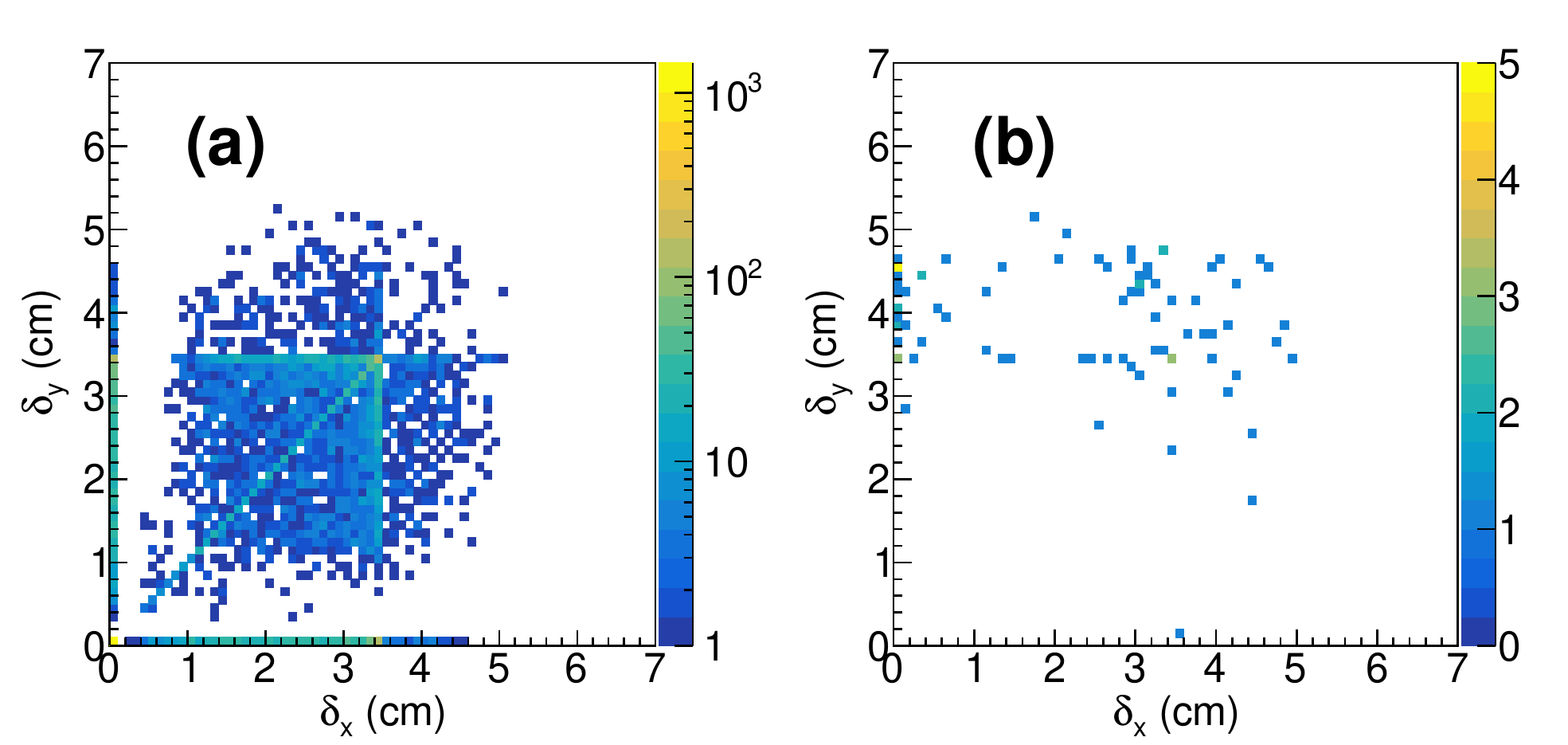}
    \caption{(Color online) Scattering plot of $\delta y$ vs.  $\delta x$ for the events with the energy (a) $E_{\rm tot}\le 80$ MeV and (b) $E_{\rm tot}> 80$ MeV}
    \label{fig:delta}
\end{figure}

Fig. \ref{fig:delta}  presents the scattering plot of $\delta y$ vs.  $\delta x$ for the events with the energy $E_{\rm tot}\le 80$ MeV (a) and $E_{\rm tot}> 80$ MeV (b), respectively. Different features can be seen in  the two groups of events. For $E_{\rm tot}\le 80$ MeV, the distribution shows nearly x-y symmetry, while for $E_{\rm tot}>80$ MeV, the distribution is obviously dominated by the events with $\delta y$ situating at a high level, consistent with the picture that the cosmic ray muons from upper penetrates through the hodoscope. 

\begin{figure} [h]
    \centering
    \includegraphics[width=.8\linewidth]{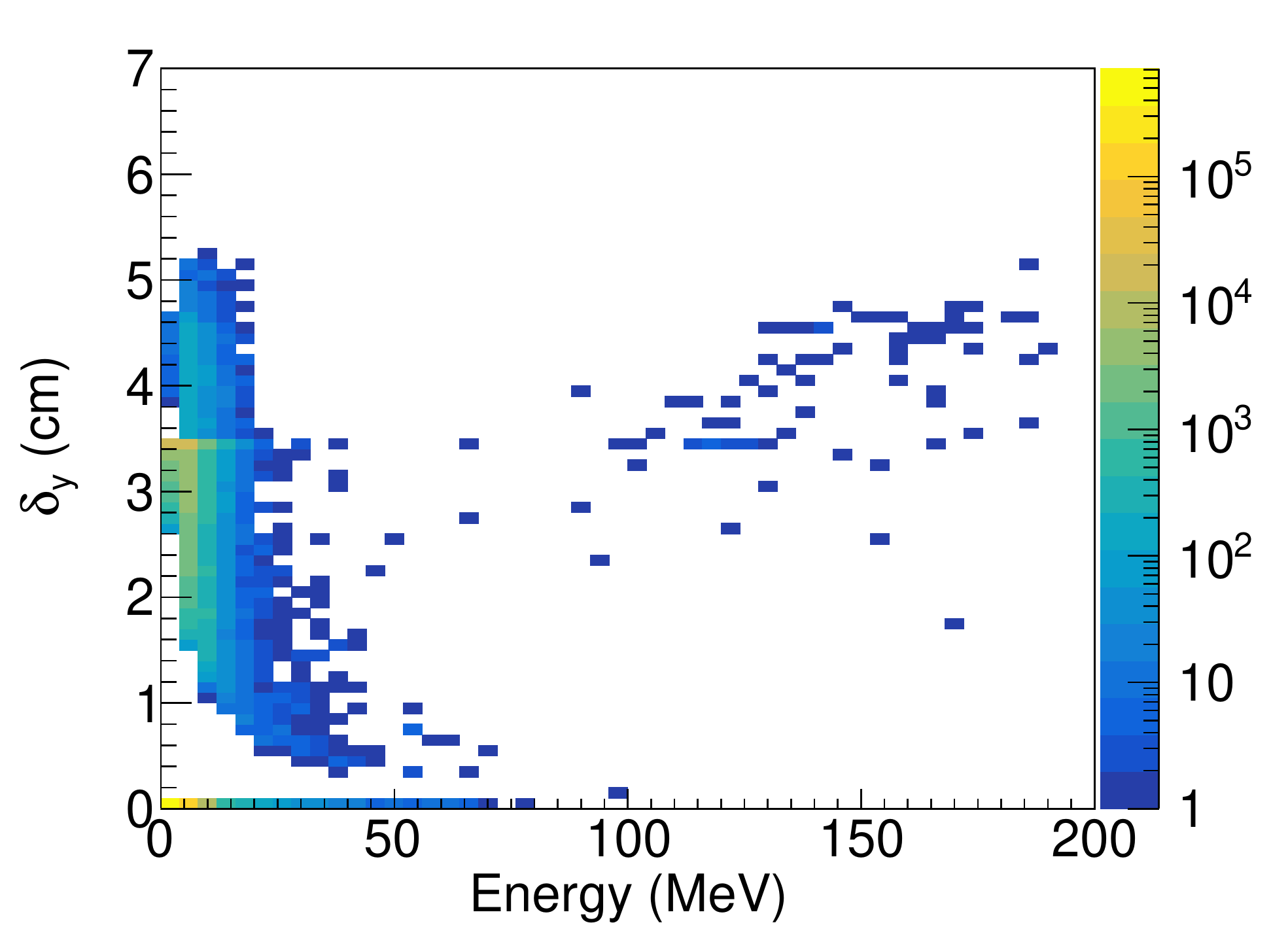}
    \caption{ (Color online)  Correlation between the spatial spreading and the total energy.}
    \label{fig:delta-E}
\end{figure}

Fig. \ref{fig:delta-E}  further presents the correlation between the total energy $E_{\rm tot}$ and the vertical spatial spreading   $\delta y$. It is shown that the events are separated in two groups, with a vacant separation at about 80 MeV. The high energy group beyond 80 MeV is the cosmic ray muons randomly correlated with the reactions within the coincident window.  The events of   $E_{\rm tot}\le 80$ MeV, are the $\gamma$ rays produced by the reactions of \krsn ~at 25 MeV/u.

Table \ref{tab:cuts} summarizes all the cuts used in the  $\gamma$ analysis.

\begin{table}[h]\centering
    \caption{Cuts used in the $\gamma$ analysis.}\label{tab:cuts}
    \begin{tabular}{ c c c}
        \hline\hline
        Cut condition & Cut values & Number of events\\ 
        \hline
         Fission trigger & $M_{\rm ppac} =2$ & 11848075\\ 
         $\gamma$ multiplicity & $\ge 1$ & 862164 \\ 
         Best Events & $I_{\rm ec}= 5, 6, 9, 10$   & 251781 \\ 
         $\gamma$ energy  & $10 \le  E_{\rm tot} \le 80$ MeV & 2159 \\
        \hline
    \end{tabular}
\end{table}

\begin{figure}[h]
    \centering
    \includegraphics[width=.9\linewidth]{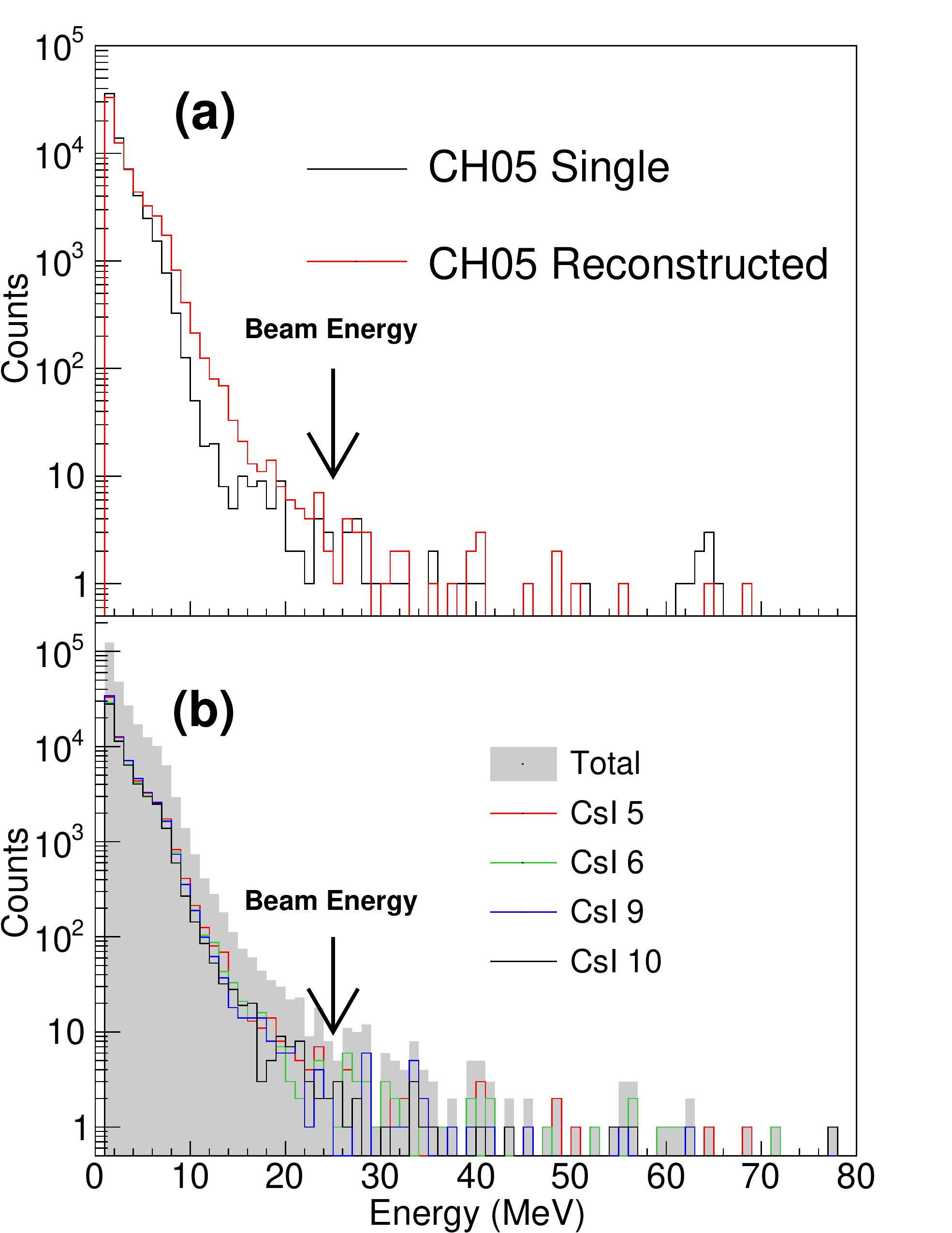}
    \caption{(Color online) Detected energy spectrum. (a) The spectra before and after reconstruction with No. 5 as the event center.  (b) The reconstructed spectra in No. 5, 6, 8 and 10 (open histograms) and the sum spectrum (filled histogram).}
    \label{fig:c_central_recon}
    \label{fig:h_central_jiugong}
\end{figure}

Figure \ref{fig:h_central_jiugong} shows the detected $\gamma$ energy spectra using the reconstruction method and cuts described above. Panel (a) presents the energy spectrum of CsI(Tl) No. 5 as event center  with (red) and without (black) adding the energy in the surrounding  crystals. It is shown that the reconstructed spectrum in the $3\times 3$ array is harder than the original spectrum in the single unit. Panel (b) presents the reconstructed spectra using No. 5, 6, 9 and 10 as the event centers, respectively. The shadowed histogram represents the total energy spectrum. The beam energy per nucleon $E_{\rm beam}$ is indicated by the arrow. It is evident that high energy $\gamma$ rays up to 3 time of $E_{\rm beam}$  is created by the bremsstrahlung radiation.  

\section{Summary}
A  hodoscope consisting of 15 CsI(Tl) crystals has been assembled  and mounted at the compact spectrometer of heavy ion experiments (CSHINE) to  measure the Bremsstrahlung high energy $\gamma$ rays from heavy ion reactions at Fermi energies.  The performance of the hodoscope has been demonstrated by the test using radioactive source and high energy $\gamma$ rays from proton induced reactions. It is suggested that the detection and reconstruction efficiency keeps constantly at $2\times 10^{-4}$ for $\gamma$ rays beyond 10 MeV. The energy resolution of single unit with $2\%+1.8\%/E_{\gamma}^{1/2}$ is obtained.  The hodoscope  has been operated in the beam experiment of 25 MeV/u $^{86}$Kr+$^{124}$Sn.  The $\gamma$ rays with the energy much higher than the beam energy per nucleon are observed. The experimental results call for further analysis and theoretic calculations of the production of  Bremsstrahlung $\gamma$ rays in heavy ion reactions.

\section* {Declaration of competing interest}
  The authors declare that they have no known competing financial interests or personal relationships that could have appeared to influence the work reported in this paper.
  
\section* {Acknowledgements}
  This work is supported by the National Natural Science Foundation of China under Grant Nos. 11875174,  11961141004,  
  by the Ministry of Science and Technology under Grant Nos. 2020YFE0202001 and 2022YFE0103400, and by Tsinghua University Initiative Scientific Research Program and the Heavy Ion Research Facility at Lanzhou (HIRFL).
  


\begin{thebibliography}{00}

\bibitem{Abbott17}
\href{https://doi.org/10.1103/PhysRevLett.119.161101}
{B.P. Abbott, et al., Phys. Rev. Lett. 119 (2017) 161101.}

\bibitem{Abbott18}
\href{https://doi.org/10.1103/PhysRevLett.121.161101}
{B.P. Abbott, et al., Phys. Rev. Lett. 121 (2018) 161101.}

\bibitem{De18}
\href{https://doi.org/10.1103/PhysRevLett.121.091102}
{S. De, D. Finstad, J. M. Lattimer et al. Phys. Rev. Lett., 121, 091102 (2018).}

\bibitem{NBZhang2018}
\href{https://doi.org/10.3847/1538-4357/aac027}
{N. B. Zhang, B. A. Li and J. Xu, Astro. Phys. J. 859, 90 (2018).}

\bibitem{NBZhang2019}
\href{https://doi.org/10.1140/epja/i2019-12700-0}
{N. B. Zhang and B. A. Li, Eur. Phys. J. A 55, 39 (2019).}

\bibitem{Xie2019}
\href{https://doi.org/10.3847/1538-4357/ab3f37}
{W. J. Xie and B. A. Li, The Astrophysical Journal, 883,174, (2019).}

\bibitem{ALi2018}
\href{https://doi.org/10.3847/1538-4357/aacc28}
{Z. Y. Zhu, E. P. Zhou, A. Li Astro. Phys. J. 862, 98 (2018).}

\bibitem{tsa04}
\href{https://doi.org/10.1103/PhysRevLett.92.062701}
{M.B. Tsang et al, Phys. Rev. Lett. 92, 062701 (2004).}

\bibitem{lba04}
\href{https://doi.org/10.1016/j.nuclphysa.2004.02.016}
{B. A. Li et al, Nuclear Physics A 735, 563 (2004).}

\bibitem{clw05}
\href{https://doi.org/10.1103/PhysRevLett.94.032701}
{L. W. Chen et al., Phys. Rev. Lett. 94, 032701 (2005).}

\bibitem{tsa09}
\href{https://doi.org/10.1103/PhysRevLett.102.122701}
{M. B. Tsang et al, Phys. Rev. Lett. 102, 122701 (2009).}

\bibitem{fam06}
\href{https://doi.org/10.1103/PhysRevLett.97.052701}
{M. A. Famiano et al., Phys. Rev. Lett. 97, 052701 (2006).}

\bibitem{Tamii2011}
\href{https://doi.org/10.1103/PhysRevLett.107.062502}
{A. Tamii et al., Phys. Rev. Lett. 107, 062502 (2011).} 


\bibitem{ZZ2015}
\href{https://doi.org/10.1103/PhysRevC.93.034335}
{Z. Zhang and L. W. Chen, Phys. Rev.  C 92, 031301(R) (2015).}

\bibitem{zhangy07}
\href{https://doi.org/10.1103/PhysRevC.95.041602}
{Y. Zhang et al., Phys. Rev. C 95, 041602(R) (2017).}

\bibitem{kol10}
\href{https://doi.org/10.1103/PhysRevC.82.064601}
{Z. Kohley et al., Phys. Rev. C 82, 064601, (2010).}

\bibitem{kol13}
\href{https://doi.org/10.1103/PhysRevC.88.041601}
{Z. Kohley et al., Phys. Rev. C 88, 041601(R), (2013).}

\bibitem{nat10}
\href{https://doi.org/10.1103/PhysRevLett.104.202501}
{J. B. Natowitz et al., Phys. Rev. Lett 104, 202501 (2010).}

\bibitem{kow07}
\href{https://doi.org/10.1103/PhysRevC.75.014601}
{S. Kowaski et al., Phys. Rev. C 75, 014601 (2007).}

\bibitem{ou15}
\href{https://doi.org/10.1103/PhysRevLett.115.212501}
{L. Ou et al., Phys. Rev. Lett. 115, 212501 (2015).}

\bibitem{ou20}
\href{https://doi.org/10.1103/PhysRevC.101.024603}
{X. Liang et al., Phys. Rev. C 101, 024603 (2020).}

\bibitem{wuqh20}
\href{https://doi.org/10.1016/j.physletb.2020.135865}
{Q. H. Wu et al., Phys. Lett. B 811, 135865(2020).}

\bibitem{yong22}
\href{https://doi.org/10.1103/PhysRevC.106.024902}
{G.C.Yong et al., Phys. Rev. C 106,024902(2022).}

\bibitem{yongjiawang15}
\href{https://doi.org/10.1140/epja/i2015-15037-8}
{Yongjia Wang et al., Eur. Phys. J. A 51, 37 (2015).}

\bibitem{yongjiawang20}
\href{https://doi.org/10.1007/s11467-020-0964-6}
{Yongjia Wang et al., Front. Phys. 15, 44302 (2020).}

\bibitem{spirit21}
\href{https://doi.org/10.1103/PhysRevLett.126.162701}
{J. Estee et al., Phys. Rev. Lett. 126, 162701 (2021).}

\bibitem{nature22}
\href{https://doi.org/10.1038/s41586-022-04750-w}
{S. Huth, et al. Nature 606, 276–280 (2022). }

\bibitem{CSHINE-Trigger}
\href{https://doi.org/10.1007/s41365-022-01149-0} 
{Dong Guo et al., Nucl. Sci. Tech. 33,  162 (2022)}

\bibitem{NIMA-CSHINE}
\href{https://doi.org/10.1016/j.nima.2021.165592}
{F. Guan et al., Nucl. Inst. Meth. A 1011, 165592 (2021).}

\bibitem{NST-CSHINE}
\href{https://doi.org/10.1007/s41365-020-00842-2}
{Y. J. Wang et al., Nucl. Sci. Tech. 32, 4 (2021).}

\bibitem{NST-Diao}
\href{https://doi.org/10.1007/s41365-022-01024-y}
{X. Y. Diao et al., Nucl. Sci. Tech. 33, 40 (2022)}

\bibitem{PLB-WYJ}
\href{https://doi.org/10.1016/j.physletb.2021.136856}
{Y. J. Wang et al., Phys. Lett. B. 825, 136856 (2022)}

\bibitem{Geant4}
\href{https://doi.org/10.1016/S0168-9002(03)01368-8}
{S. Agostinelli et al., Nucl. Inst. Meth. A 506(3), 250 (2003).}

\bibitem{CIAE1} 
\href{https://doi.org/10.1088/1674-1056/28/6/060706}
{Y. L. Dang et al., Chin. Phys. B 28(6), 060706 (2019).}

\bibitem{CIAE2}
\href{http://cpb.iphy.ac.cn/CN/10.1088/1674-1056/ab96a0}
{F. L. Liu et al., Chin. Phys. B 29(7), 070702 (2020).}

\bibitem{CIAE3}
\href{https://doi.org/10.1016/j.nimb.2022.08.005}
{F. L. Liu et al., Nucl. Inst. Meth. B 529, 56 (2022)}

\bibitem{BESIII-const}
\href{https://doi.org/10.1016/j.nima.2008.08.072}
{BESIII Collaboration, Nucl. Inst. Meth. A 598(1), 7 (2008)}

\bibitem{BaBar-EMC}
\href{https://doi.org/10.1016/S0168-9002(02)01484-5}
{B. Lewandowski, Nucl. Inst. Meth. A 494(1), 303 (2002)}

\bibitem{TAPS-IEEE}
\href{https://doi.org/10.1109/23.289329}
{R. Novotny, IEEE Trans. Nucl. Sc. 38(2), 379 (1991)}

\bibitem{TAPS-NIMA}
\href{https://doi.org/10.1016/0168-9002(95)00525-0}
{F. M. Marqu\'es et al., Nucl. Inst. Meth. A 365(2), 392 (1995)}

\end{thebibliography}
\end{document}